\documentclass[twocolumn]{aastex631}

\usepackage{graphicx}	
\usepackage{amsmath}	
\usepackage{amssymb}	
\usepackage{ae,aecompl}

\shorttitle{WASP-47: ESPRESSO \& TESS}
\shortauthors{Bryant et al.}

\newcommand{\hjs}{hot Jupiters}
\newcommand{\hj}{hot Jupiter}
\newcommand{\wjs}{warm Jupiters}
\newcommand{\wj}{warm Jupiter}

\newcommand{\mpl}{\mbox{M$_{P}$}}
\newcommand{\rpl}{\mbox{R$_{P}$}}
\newcommand{\rhopl}{\mbox{$\rho_{P}$}}
\newcommand{\mstar}{\mbox{M$_{*}$}}
\newcommand{\rstar}{\mbox{R$_{*}$}}
\newcommand{\rhostar}{\mbox{$\rho_{*}$}}
\newcommand{\mjup}{\mbox{M$_{J}$}}
\newcommand{\rjup}{\mbox{R$_{J}$}}

\newcommand{\mearth}{\mbox{M$_{\oplus}$}}
\newcommand{\rearth}{\mbox{R$_{\oplus}$}}
\newcommand{\msun}{\mbox{M$_{\odot}$}}
\newcommand{\rsun}{\mbox{R$_{\odot}$}}
\newcommand{\gccc}{g\,cm$^{-3}$}

\newcommand{\teff}{$T_{\rm eff}$}
\newcommand{\feh}{\mbox{[Fe/H]}}

\newcommand{\logg}{$\log g$}
\newcommand{\tc}{$T_C$}

\newcommand{\prot}{$P_{\rm Rot}$}
\newcommand{\protmin}{$P_{\rm Rot; \ min}$}
\newcommand{\cnce}{55-Cancri\,e}
\newcommand{\cnc}{55-Cancri}

\newcommand{\logrhk}{$\log R^{\prime}_{\rm HK}$}
\newcommand{\Nstar}{WASP-47}
\newcommand{\NRA}{\mbox{$331.20309414396$}} 
\newcommand{\NDec}{\mbox{$-12.01907283535$}} 
\newcommand{\NpropRA}{\mbox{$15.074\pm0.020 $}} 
\newcommand{\NpropDec}{\mbox{$-41.467\pm0.020$}} 

\newcommand{\Nparallax}{\mbox{$3.7010\pm0.0201$}}

\newcommand{\Nmetal}{\mbox{$0.38\pm0.05$}} 
\newcommand{\Nlogg}{\mbox{$4.3437  \pm 0.0063$}}

\newcommand{\Nprotmin}{$28.762$}
\newcommand{\Nmetalcnc}{\mbox{$0.35\pm0.10$}}
\newcommand{\Ngamesp}{\mbox{$-27165.95 \pm 1.27$}}
\newcommand{\Ngamfwhm}{\mbox{$7635.6 \pm 1.0$}}
\newcommand{\Ngamharps}{\mbox{$-27041.03 \pm 0.55$}}
\newcommand{\Ngamhires}{\mbox{$6.50 \pm 0.75$}}
\newcommand{\NgamcorA}{\mbox{$-27081.33 \pm 1.89$}}
\newcommand{\NgamcorB}{\mbox{$-27067.57 \pm 5.96$}}
\newcommand{\Nlogjitesp}{\mbox{$0.145\pm 0.214$}}
\newcommand{\Nlogjitharps}{\mbox{$-2.33 \pm 1.82$}}
\newcommand{\Nlogjithires}{\mbox{$0.488 \pm 0.069$}}
\newcommand{\NlogjitcorA}{\mbox{$-2.38 \pm 3.03$}}
\newcommand{\NlogjitcorB}{\mbox{$-2.95 \pm 2.75$}}
\newcommand{\Nlogjitfwhm}{\mbox{$-2.79 \pm 0.13$}}
\newcommand{\Nprot}{\mbox{$39.4\,^{+2.2}_{-4.5}$}}
\newcommand{\Nlogq}{\mbox{$2.6\,^{+5.2}_{-1.1}$}}
\newcommand{\Nlogdq}{\mbox{$-0.40\,^{+6.18}_{-5.04}$}}
\newcommand{\Nf}{\mbox{$0.52\,^{+0.32}_{-0.31}$}}
\newcommand{\Nsiggpesp}{\mbox{$4.58\,^{+3.83}_{-2.44}$}}
\newcommand{\Nsiggpfwhm}{\mbox{$7.04\,^{+4.18}_{-2.16}$}}
\newcommand{\NVmag}{$11.936 \pm 0.046$}
\newcommand{\NBmag}{$12.736 \pm 0.024$}

\newcommand{\NTmag}{$11.29 \pm 0.0061$}
\newcommand{\NGAIAmag}{$11.7817 \pm 0.0028$}
\newcommand{\Ngaiabp}{$12.1716 \pm 0.0029$}
\newcommand{\Ngaiarp}{$11.2294 \pm 0.0038$}

\newcommand{\NJmag}{$10.613 \pm 0.022$}
\newcommand{\NHmag}{$10.310 \pm 0.022$}
\newcommand{\NKmag}{$10.192 \pm 0.026$}

\newcommand{\Nplanete}{WASP-47\,e}
\newcommand{\Nperiode}{\mbox{$0.7895933 \pm 0.0000044$}}
\newcommand{\Nperiodshorte}{\mbox{$0.7896$}}
\newcommand{\Ntce}{\mbox{$2457011.34862 \pm 0.00030$}}
\newcommand{\Nmasse}{\mbox{$6.77\pm 0.57$}}%
\newcommand{\NKe}{\mbox{$4.55\pm0.37$}}
\newcommand{\Ntsme}{\mbox{$13.2\pm1.3$}}
\newcommand{\Nradiuse}{\mbox{$1.808\pm 0.026$}}%
\newcommand{\Ndensitye}{\mbox{$6.29\pm0.60$}}%
\newcommand{\Nrratioe}{\mbox{$0.01458\pm0.00013$}}

\newcommand{\Nplanetb}{WASP-47\,b}
\newcommand{\Nperiodb}{\mbox{$4.1591492 \pm 0.0000006$}}
\newcommand{\Nperiodshortb}{\mbox{$4.159$}}
\newcommand{\Ntcb}{\mbox{$2457007.932103 \pm 0.000019$}}
\newcommand{\Nmassb}{\mbox{$363.6\pm7.3$}}%
\newcommand{\Nmassbjup}{\mbox{$1.144\pm0.023$}}%
\newcommand{\NKb}{\mbox{$140.84\pm0.40$}}
\newcommand{\Ntsmb}{\mbox{$46.8\pm2.3$}}
\newcommand{\Nradiusb}{\mbox{$12.64\pm0.15$}}%
\newcommand{\Ndensityb}{\mbox{$0.989\pm0.040$}}%
\newcommand{\Nrratiob}{\mbox{$0.10191\pm0.00022$}}

\newcommand{\Nplanetc}{WASP-47\,c}
\newcommand{\Nperiodc}{\mbox{$588.8 \pm 2.0$}}
\newcommand{\Nperiodshortc}{\mbox{$588.4$}}
\newcommand{\Ntcc}{\mbox{$2457763.1 \pm 4.3$}}
\newcommand{\Neccc}{\mbox{$0.295 \pm 0.016$}}%
\newcommand{\Nomegac}{\mbox{$112.0 \pm 4.3$}}
\newcommand{\Nmassc}{\mbox{$398.9\pm9.1$}}%

\newcommand{\NKc}{\mbox{$31.04\pm0.40$}}

\newcommand{\Nplanetd}{WASP-47\,d}
\newcommand{\Nperiodd}{\mbox{$9.03055 \pm 0.00008$}}
\newcommand{\Nperiodshortd}{\mbox{$9.031$}}
\newcommand{\Ntcd}{\mbox{$2457006.36955 \pm 0.00035$}}
\newcommand{\Neccd}{\mbox{$0.010\,^{+0.011}_{-0.007}$}}   
\newcommand{\Nomegad}{\mbox{$16.5\,^{+84.2}_{-98.6}$}}
\newcommand{\Nmassd}{\mbox{$14.2\pm1.3$}}%
\newcommand{\NKd}{\mbox{$4.26\pm0.37$}}
\newcommand{\Ntsmd}{\mbox{$23.4\pm2.4$}}
\newcommand{\Nradiusd}{\mbox{$3.567\pm0.045$}}%
\newcommand{\Ndensityd}{\mbox{$1.72\pm0.17$}}%
\newcommand{\Nrratiod}{\mbox{$0.02876\pm0.00017$}}

\usepackage{newtxtext,newtxmath}

\begin{document}

\title{Revisiting WASP-47 with ESPRESSO and TESS}

\author[0000-0002-0786-7307]{Edward M. Bryant}
\affiliation{Dept.\ of Physics, University of Warwick, Gibbet Hill Road, Coventry CV4 7AL, UK}
\affiliation{Centre for Exoplanets and Habitability, University of Warwick, Gibbet Hill Road, Coventry CV4 7AL, UK}

\author[0000-0001-6023-1335]{Daniel Bayliss}
\affiliation{Dept.\ of Physics, University of Warwick, Gibbet Hill Road, Coventry CV4 7AL, UK}
\affiliation{Centre for Exoplanets and Habitability, University of Warwick, Gibbet Hill Road, Coventry CV4 7AL, UK}

\begin{abstract}
\Nstar\ hosts a remarkable planetary system containing a \hj\ (\Nplanetb; P = \Nperiodshortb\,days) with an inner super-Earth (\Nplanete; P = \Nperiodshorte\,days), a close-orbiting outer Neptune (\Nplanetd; P = \Nperiodshortd\,days), and a long period giant planet (\Nplanetc; P = \Nperiodshortc\,days). We use the new TESS photometry to refine the orbital ephemerides of the transiting planets in the system, particularly the \hj\ \Nplanetb, for which we find an update equating to a 17.4\,min shift in the transit time. We report new radial velocity measurements from the ESPRESSO spectrograph for \Nstar, which we use to refine the masses of \Nplanetd\ and \Nplanete, with a high cadence observing strategy aimed to focus on the super-Earth \Nplanete. We detect a periodic modulation in the K2 photometry that corresponds to a 32.5$\pm$3.9\,day stellar rotation, and find further stellar activity signals in our ESPRESSO data consistent with this rotation period. For \Nplanete\ we measure a mass of \Nmasse\,\mearth\ and a bulk density of \Ndensitye\,\gccc, giving \Nplanete\ the second most precisely measured density to date of any super-Earth.  The mass and radius of \Nplanete, combined with the exotic configuration of the planetary system, suggest the \Nstar\ system formed through a mechanism different to systems with multiple small planets or more typical isolated \hjs.
\end{abstract}




\section{Introduction}\label{intro}

The formation mechanisms for \hj\ planets ($\rpl > 0.6\,\rjup$; $P < 10$\,days) remain uncertain, although viable formation pathways have been proposed \citep{dawsonjohnson2018hotjupiters}. One possible scenario is that \hjs\ begin their formation at large orbital distances and then migrate towards the star during formation \citep[eg.][]{lin199651pegmigration, nelson2000migration}. This planetary migration could arise from multiple causes, such as interactions between the protoplanet and the circumstellar disk \citep{ppaploizou2000discmigration} or through some form of high eccentricity migration, which can be triggered by planet-planet scattering \citep{chatterjee2008planetscattering} or secular interactions between multiple bodies, including Kozai-Lidov cycles \citep{kozai1962, lidov1962}. This high eccentricity migration removes inner planets from the system \citep{mustill2015hem}, leaving most \hj\ planets isolated and without other planetary companions in close orbits \citep[eg.][]{steffen2012}.  The observational evidence seems to support this model.  The vast majority of \hjs\ do not have close orbiting companions, although many have long-period massive outer companions \citep[eg.][]{knutson2014hjfriendsI}.  For the longer period \wjs\ the situation is different, and \wj\ planets are significantly more likely to be accompanied by closely orbiting small planetary companions \citep{huang2016lonelyHJs}.  A possible explanation for this is that these \wjs\ are forming in-situ, as opposed to migrating from outer regions.

Out of the hundreds of \hjs\ discovered to date, only three have been found to also host small, inner planets: Kepler-730c \citep{canas2019kep730}, TOI-1130c \citep{huang2020toi1130}, and \Nplanetb\ \citep{hellier2012wasp47}. The \cnc\ system \citep{bourrier201855cnce} is also similar to the \Nstar\ system in that it contains a super-Earth with an orbital period shorter than one day (\cnce), and close--in giant planet (55-Cancri b; 14.6\,days). As a result of the brightness of the host star (V = 5.95\,mag), the \cnc\ system, and particularly \cnce, have been very well studied \citep[eg.][]{demory201255cncemission, tsiaras201655cancrihstatmos, angelo201755cancrispitzeratmos}.

Since the configurations of these three systems bears more resemblance to the population of \wjs\ than to the \hjs\, it has been speculated that perhaps these systems also formed through an in-situ pathway \citep{huang2020toi1130}. It is therefore imperative to study these three systems in detail to determine if there is any other evidence that they formed from a different pathway to the majority of \hj\ systems.  In order to do this, we require precise knowledge of the planetary system parameters, especially the planetary masses, densities, and orbital periods.

\begin{figure}
    \centering
    \includegraphics[width=\columnwidth]{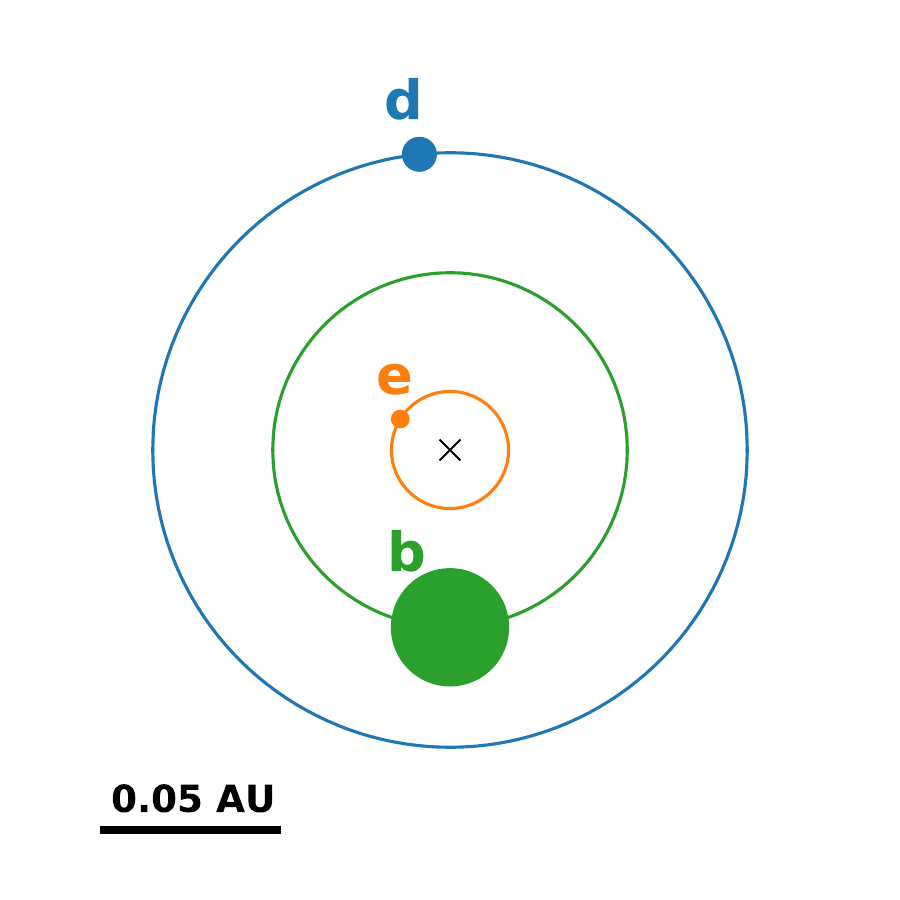}
    \caption{Orbital configuration of the inner three planets in the  \Nstar\ system. The three lines show the orbits of the planets: \Nplanetb\ (green), \Nplanetd\ (blue), \Nplanete\ (orange). The position of the \Nstar\ star is depicted by the black cross.  The filled circles give the positions of the planets at T = 2549449.35 BJD, which is the time of the first transit of \Nplanetb\ observed by TESS, assuming observer viewing from bottom of plot and orbits in the anti-clockwise direction. The relative sizes of the filled circles are proportional to the the relative radii of the planets.}
    \label{fig:orb_config}
\end{figure}

\Nstar\ has been extensively studied since its discovery in 2012.  Initially, \Nstar\ was shown to to host a \Nperiodshortb\,day period transiting hot-Jupiter \Nplanetb\ \citep{hellier2012wasp47}. Later, an extensive spectroscopic monitoring campaign revealed the presence of the \Nperiodshortc\,day giant planet \Nplanetc\ \citep{neveuVanMalle2016wasp47Coralie}. It is not currently known whether or not \Nplanetc\ transits the star, although \citet{vanderburg2017wasp47} derived a 10\% probability that it does transit. This estimate is greater than the geometric probability ($\approx$ 0.6\%) and arises from additional dynamical constraints placed by the stability of the inner \Nstar\ system on the inclination of \Nplanetc.  Most surprisingly, two additional small planets were discovered through photometric monitoring \citep{becker2015wasp47K2} with the Kepler space telescope during the K2 mission \citep{howell2014K2}. \Nplanetd\ is a Neptune-mass planet exterior to the \hj, while  \Nplanete\ is a super-Earth interior to the \hj. The orbital configuration of the inner regions of this remarkable planetary system are shown in Figure~\ref{fig:orb_config}.

\begin{table}
	\centering
	\caption{Key Stellar Properties for \Nstar}
	\begin{tabular}{lcc} 
	Property	&	Value		&Source\\
	\hline
    \multicolumn{3}{l}{}\\
    TIC & 102264230 & TICv8 \\
    \\
    RA (deg)  & \NRA & \textit{Gaia} EDR3 \\
    Dec (deg) & \NDec & \textit{Gaia} EDR3 \\
    $\mu_{\rm RA}$ (mas) & \NpropRA & \textit{Gaia} EDR3 \\
    $\mu_{\rm Dec}$ (mas) & \NpropDec & \textit{Gaia} EDR3 \\
    Parallax (mas) & \Nparallax & \textit{Gaia} EDR3 \\
    \\
    V (mag)		&\NVmag 	& TICv8 \\
	B (mag)		&\NBmag		& TICv8 \\
	TESS (mag)  &\NTmag     & TICv8 \\
	\textit{Gaia} G (mag) & \NGAIAmag & \textit{Gaia} EDR3 \\
	\textit{Gaia} b$_{\rm p}$ (mag) & \Ngaiabp & \textit{Gaia} EDR3 \\
	\textit{Gaia} r$_{\rm p}$ (mag) & \Ngaiarp & \textit{Gaia} EDR3 \\
	J (mag) & \NJmag & 2MASS \\
	H (mag) & \NHmag & 2MASS \\
	K (mag) & \NKmag & 2MASS \\
	\\
	\mstar\ (\msun) & 1.040 $\pm$ 0.031 & V2017 \\
	\rstar\ (\rsun) & 1.137 $\pm$ 0.013 & V2017 \\
	\rhostar\ (\gccc) & 0.998 $\pm$ 0.014 & Section~\ref{sec:trans_analysis}\\
	\teff\ (K) & 5552 $\pm$ 75 & V2017 \\
	\feh & 0.38 $\pm$ 0.05 & V2017 \\
	\logg (cgs) & \Nlogg & V2017 \\
    \\
    \hline
    \multicolumn{3}{l}{TICv8 - \citet{stassun2019ticv8}} \\
    \multicolumn{3}{l}{\textit{Gaia} EDR3 - \citet{gaia2021edr3}} \\
    \multicolumn{3}{l}{2MASS - \citet{skrutskie2006twomass}} \\
    \multicolumn{3}{l}{V2017 - \citet{vanderburg2017wasp47}} \\
    \\
    \hline
	\end{tabular}
	\label{tab:stellar}
\end{table}

Further to these discovery papers, there have been many independent efforts to measure the masses of the planets in the \Nstar\ system. Multiple radial velocity monitoring campaigns using the PFS \citep{dai2015wasp47PFS}, HIRES \citep{sinukoff2017wasp47hires}, and HARPS-N \citep{vanderburg2017wasp47} spectrographs have been carried out. Dynamical analyses have also been performed to determine the masses of the planets \citep[eg.][]{almenara2016wasp47_dynmass, weiss2017wasp47rvttv}. These different studies reached slightly differing conclusions about the mass, and therefore the composition, of \Nplanete.

In addition to the consequences for planetary formation, \Nplanete\ represents one of the best cases to study the composition of a super-Earth like planet, thanks to the wealth of data available on the system. Therefore, we seek to shed yet more light on this system. 

In this work we use the next generation of high precision spectrographs, ESPRESSO, in order to obtain the most precise and accurate measurements of the planet masses and densities, particularly focusing on the super-Earth \Nplanete.

\section{Observations}\label{sec:obs}
\subsection{ESPRESSO}\label{sec:obs:espresso}
ESPRESSO \citep[Echelle SPectrograph for Rocky Exoplanets and Stable Spectroscopic Observations;][]{pepe2020espresso} is a new high-resolution, visible spectrograph operating at the VLT at ESO's Paranal Observatory in Chile. ESPRESSO can be fed from any of the 8.2\,m Unit Telescopes and can also be fed simultaneously by all four. The three main exoplanet science objectives of ESPRESSO are to find new Earth-mass planets orbiting in the habitable zone of Sun-like stars, characterize the atmospheres of exoplanets, and precisely determine the masses of low-mass transiting exoplanets. ESPRESSO has already been used to confirm TESS discoveries, including  LP 714-47\,b \citep[TOI-442 b; ][]{dreizler2020lp71447}, TOI-130\,b \citep{sozzetti2021toi130}, and the planets in the TOI-178 system \citep{leleu2021toi178}. ESPRESSO radial velocity measurements have also been used to improve upon the precision of the masses of LHS-1140\,b and c \citep{lillobox2020lhs1140}.

ESPRESSO operates in the wavelength range 380--788\,nm, and is designed to achieve radial velocity precision of 10\,cm\,s$^{-1}$ for bright stars (V < 8\,mag) in order to be capable of detecting Earth-mass planets around solar type stars.  This unprecedented radial velocity precision is achieved both by building on and improving the technologies used in the HARPS spectrograph for stability and calibration accuracy but also through the increased light-collecting capacity of the VLT UTs compared to the ESO 3.6\,m.

\Nstar\ was observed by ESPRESSO between the dates of 2019 6 August and 30 September (Run ID: 0103.C-0422; PI Bayliss), using an exposure time of 1250\,seconds and with an airmass limit of z < 1.5 for each observation. We reduced all the spectra using the ESPRESSO reduction pipeline (version 2.2.1) through the ESOReflex workflow environment \citep{esoreflex2013}. The signal-to-noise achieved from our observations ranged from 40-60 at $\lambda$ = 550\,nm. The radial velocity CCFs were computed using a G9 mask, and the pipeline automatically extracts diagnostic information on the CCFs, including the FWHM of the CCF. In total, \Nstar\ was observed 25 times by ESPRESSO.  Two of these observations show anomalous CCF profiles likely caused by cloud coverage and/or moon contamination, and are not used in the analysis.  An additional four observations were identified as having been taken during a transit of \Nplanetb. We also remove these from our analysis so that the Rossiter-McLaughlin signal of \Nplanetb\ \citep{sanchisojeda2015wasp47RM} does not affect our radial velocity model and analysis. This left us with 19 radial velocity measurements from ESPRESSO for our analysis. We run spectral analysis on our co-added ESPRESSO spectra using the ESPRESSO-DAS pipeline. The values we derive for effective temperature and metallicity are consistent with those derived by \citet{vanderburg2017wasp47}.

In addition to the ESPRESSO data, we utilize a number of archival radial velocity measurements of \Nstar, which were obtained using the HARPS-N \citep[69 data points with mean a precision of 3.2\,ms$^{-1}$;][]{vanderburg2017wasp47}, HIRES \citep[43 data points with mean a precision of 2.0\,ms$^{-1}$;][]{sinukoff2017wasp47hires}, PFS \citep[26 data points with mean a precision of 3.2\,ms$^{-1}$;][]{dai2015wasp47PFS}, and CORALIE \citep[52 data points with mean a precision of 12.5\,ms$^{-1}$;][]{neveuVanMalle2016wasp47Coralie} spectrographs. We did not perform any additional reduction of these data sets.

\subsection{K2}\label{sec:obs:K2}
As we discuss in Section\,\ref{sec:analysis}, a degree of variability in the radial velocity data indicates relatively strong stellar activity on \Nstar.  In order to help understand and account for this activity, we turned to the high precision time series photometry for \Nstar\ from the Kepler space telescope \citep{borucki2010kepler}.

\Nstar\ was observed by the Kepler space telescope during Campaign 03 of the K2 mission \citep{howell2014K2}. \Nstar\ was observed continuously for 67\,days between the dates of 2014 November 17 and 2015 January 23. \Nstar\ was observed in the short cadence mode, and a short cadence light curve was extracted by \citet{becker2015wasp47K2} using the method of \citet{vanderburg2015hip116454}. We accessed this light curve for the transit analysis performed in this work\footnote{The short cadence light curve was accessed from \href{http://www.cfa.harvard.edu/~avanderb/wasp47sc.csv}{http://www.cfa.harvard.edu/~avanderb/wasp47sc.csv}}.

\subsection{TESS}\label{sec:obs:tess}
\begin{figure*}
    \centering
    \includegraphics[width=\textwidth]{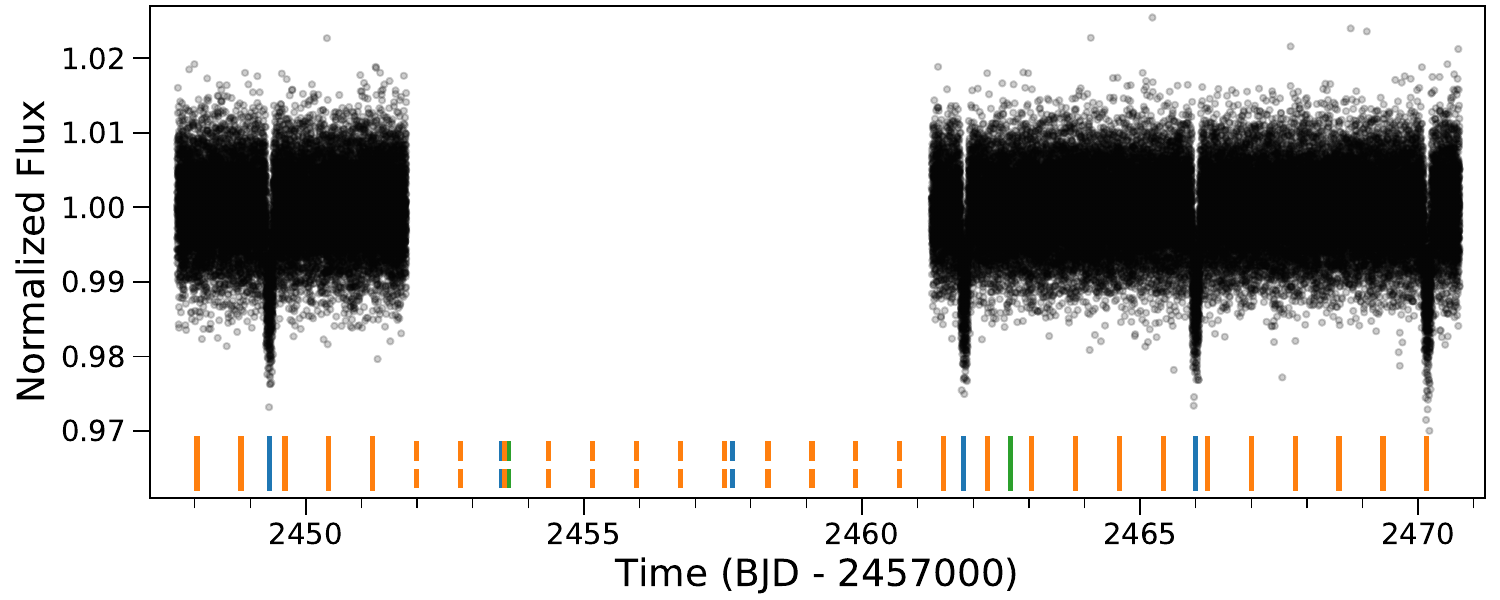}
    \caption{TESS 20\,s \textsc{PDCSAP\_FLUX} photometry for \Nstar. The colored bars at the bottom show the positions of the transits of \Nplanetb\ (blue), \Nplanetd\ (green), and \Nplanete\ (orange), with the dashed bars showing the positions of transits which fall in the large data gap caused by Earth and Moon crossing events.}
    \label{fig:tess_time}
\end{figure*}

\Nstar\ was observed by the TESS mission \citep{ricker2014tess} during Sector 42 between the dates of 2021 August 21 and September 15. \Nstar\ fell on camera 1 CCD 4. We utilized the 20\,second cadence light curve produced by the SPOC pipeline \citep{jenkins2016spoc}, which we downloaded from MAST. We use the \textsc{PDCSAP\_FLUX} time series in this work, which has had instrumental and blending effects corrected in the light curve \citep{jenkins2016spoc}.

During both orbits of Sector 42, the Earth crossed the field of view of camera 1, with the Moon also crossing the field of during the first orbit\footnote{Data release notes for Sector 42 available \href{https://archive.stsci.edu/missions/tess/doc/tess_drn/tess_sector_42_drn60_v01.pdf}{here}.}. Due to the significant increases in scattered light in the background during these events, the \textsc{PDCSAP\_FLUX} time series spans only 4.11\,days in the first orbit and 9.48\,days in the second, giving a total of around half the nominal 27\,day coverage for a given sector. The TESS light curve is shown in Figure~\ref{fig:tess_time}. \\

\section{Analysis}\label{sec:analysis}
\subsection{Transit Analysis}\label{sec:trans_analysis}
\begin{figure}
    \centering
    \includegraphics[width=\columnwidth]{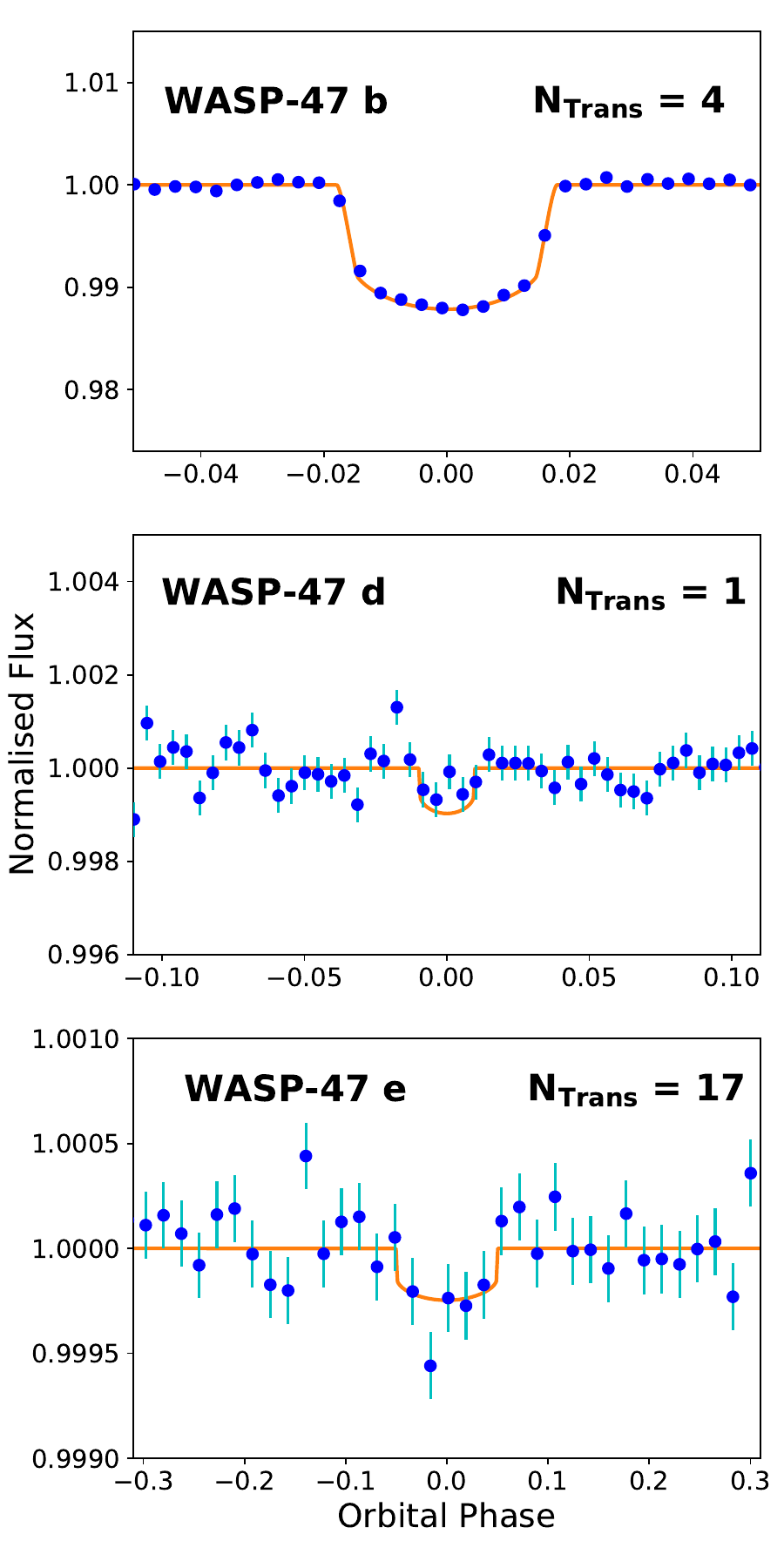}
    \caption{Phase-folded TESS photometry of \Nstar\ zoomed around the transit events for \textbf{Top:} \Nplanetb\ \textbf{Middle:} \Nplanetd\ and \textbf{Bottom:} \Nplanete. For all panels the photometry (blue circles) has had the transit models from the other two planets subtracted and is binned in phase on a time scale of 20\,minutes for \Nplanetb\ and \Nplanete\ and 60\,minutes for \Nplanetd. The orange lines show the best fit transit models from the modelling in Section~\ref{sec:trans_analysis}.}
    \label{fig:tess}
\end{figure}
\begin{figure}
    \centering
    \includegraphics[width=\columnwidth]{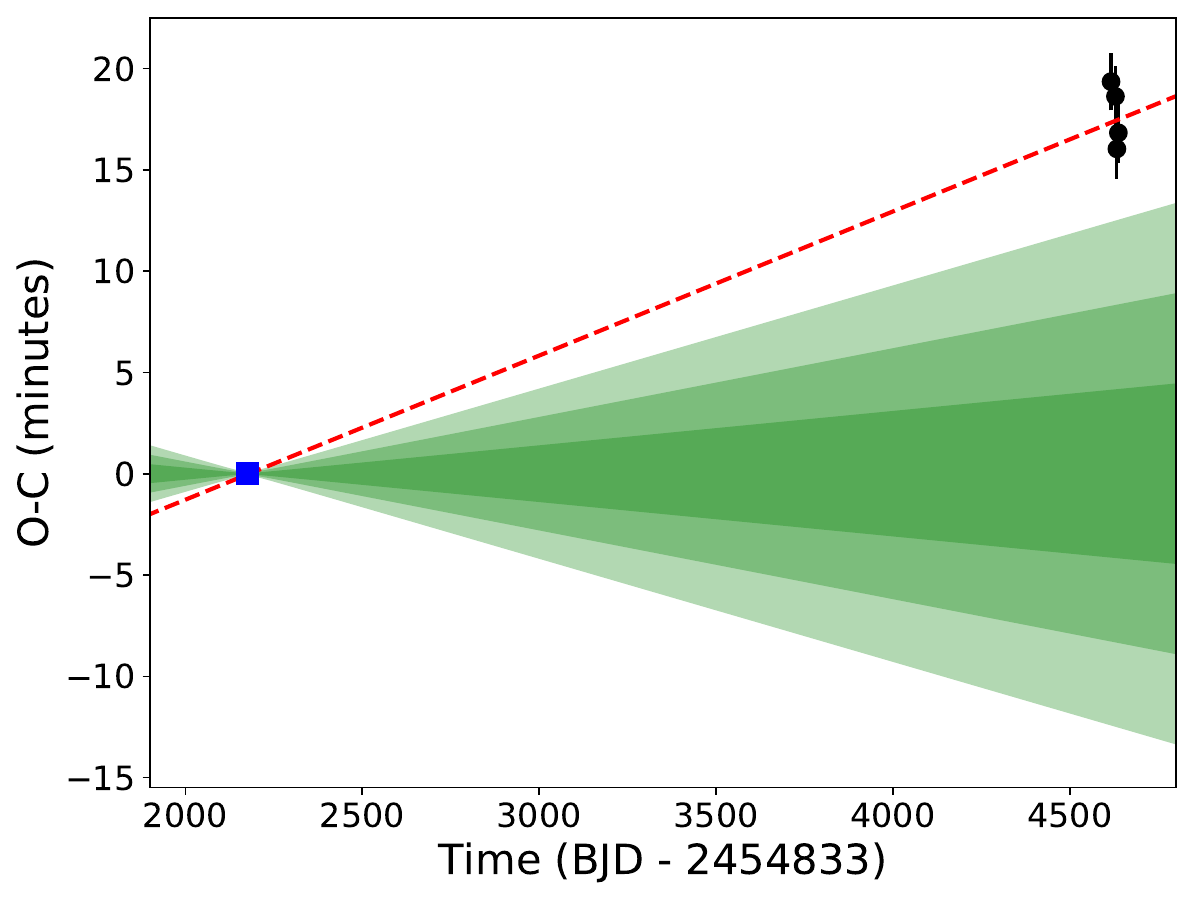}
    \caption{Measured transit times for \Nplanetb\ transit events in the TESS data (black circles) compared with the literature ephemeris (blue square and green shaded regions) from \citet{becker2015wasp47K2}.  The uncertainties on the TESS transit times have had an additional 1\,minute added in quadrature to account for the TTVs of \Nplanetb. The shaded green regions give the 1, 2, and 3\,$\sigma$ uncertainties of the predicted ephemeris from \citet{becker2015wasp47K2}. The red dashed line represents the updated ephemeris determined in this work.}
    \label{fig:w47b_ephem}
\end{figure}
We analysed the TESS data in order to refine the planetary parameters of the \Nstar\ planets. It is not known if \Nplanetc\ transits the host star \citep{neveuVanMalle2016wasp47Coralie}.  However based on the best ephemeris for \Nplanetc\, it is not expected to transit during the TESS monitoring; the next conjunction of \Nplanetc\ is predicted to occur just under 60\,days after the end of the TESS observations. We searched the light curve evidence of of any previously unknown transiting planets. We mask out the transits of \Nplanetb, d, and e, and search for additional transit signals using BLS \citep{kovacs2002bls} but we do not find any evidence for a previously unknown transiting planet. Therefore, we consider just \Nplanetb, d, and e in this analysis.

We model the transit light curves of the three planets using \textsc{batman} \citep{kreidberg2015batman} with the following free parameters: the times of transit center, $T_{C}$, the orbital periods of the planets, $P$, the planet-to-star radius ratios, $R_P / R_*$, the orbital inclinations, $i$, and the stellar density, $\rho_*$, from which we can compute the scaled semi-major axes, $a / R_*$. For $T_{C}$ and $P$ we use wide uniform priors centred on the values derived by \citet{becker2015wasp47K2}. For $R_P / R_*$ and $i$ we use uniform priors between 0 and 1 and between 0$^{\circ}$ and 90$^{\circ}$ respectively. For $\rho_*$ we use another wide uniform prior between 0.9 and 1.1, based on the prior knowledge of the stellar parameters from \citet{vanderburg2017wasp47}. We use a quadratic limb-darkening law with two independent sets of coefficients for the K2 and TESS data. We sampled for these coefficients using the parameterization of \citet{kipping2013ld}. For the analysis of \Nplanetb\ and \Nplanete\ we adopted circular orbits based on the findings of \citet{vanderburg2017wasp47} that the tidal circularization timescales for these two planets are significantly shorter than the age of the system. We allow for a non-circular orbit for \Nplanetd\ and impose a half-Gaussian prior with a width of 0.014 and centered on 0 for $e_d$. This constraint on the eccentricity of \Nplanetd\ comes from the dynamical analyses performed independently by \citet{becker2015wasp47K2} and \citet{weiss2017wasp47rvttv}. This eccentricity is taken into account when calculating the value of $a / R_*$ from \rhostar\ for the orbit of \Nplanetd. We explore the parameters using a Monte Carlo Markov Chain (MCMC) analysis using the \textsc{emcee} Ensemble Sampler \citep{foremanmackey2013emcee}. A total of 48 walkers were run for a burn in of 3000 steps followed by a further 10000 steps per chain to sample the posterior distribution. We calculated the autocorrelation lengths, $\tau$, for each parameter and find $\tau \ll N/100$, where $N$ is the chain length, indicating good convergence for all parameters except $\omega_d$ which has $\tau = N/26.5$. This is unsurprising, as due to the eccentricity of \Nplanetd\ being low and consistent with 0 at 2$\sigma$, from the transit light curves alone we cannot place strong constraints on $\omega_d$ and so do not expect excellent convergence. The transit models recovered from this analysis are plotted in Figure~\ref{fig:tess}. The planetary radii we derive are reported in Table~\ref{tab:params}. From this analysis, we obtain a stellar density of $\rho_* = 0.998 \pm 0.014$\,\gccc, which we note is consistent with the current best estimates for the stellar mass and radius \citep{vanderburg2017wasp47}.

Compared to the parameters from just the K2 photometry alone, the inclusion of the TESS photometry into the analysis does not provide additional information on the radii of the planets. This is a result of both the reduced time coverage of the TESS photometry compared to the K2 (13.59\,days vs 67\,days) and the significantly reduced photometric precision (3040\,ppm-per-minute vs 350\,ppm-per-minute). In particular, \Nplanetd\ transited just once during the available TESS photometry (see Figure~\ref{fig:tess_time}), and so the recovery of the transit in the TESS data is marginal.

TESS data have provided a useful method for refining the ephemerides for known transiting hot Jupiters by increasing the baseline of transit observations \citep{shan2021hotjupiterephems}. We are able to take advantage of this method by increasing the baseline of observations to refine the orbital periods of the \Nstar\ planets. For the hot Jupiter \Nplanetb\ our results yield an orbital period of 4.1591492$\pm$0.0000006\,days, which is significantly (4.15\,$\sigma$) longer than the current literature value \citep{becker2015wasp47K2} (see Figure~\ref{fig:w47b_ephem}). We also reduce the uncertainty on this period by a factor of seven. \Nplanetb\ displays transit timing variations with an amplitude on the order of 1\,minute \citep{becker2015wasp47K2} however these cannot account for the offset in transit times observed, which has a magnitude of 17.4\,minutes. For \Nplanetd\ we find an orbital period of 9.03055$\pm$0.00008\,days, which is slightly shorter than the \citet{becker2015wasp47K2} period, but only differs by just over 1$\sigma$.  We have reduced the uncertainty on this period by a factor of 2.4. For the super Earth \Nplanete\ our measured orbital period of 0.789595$\pm$0.000005\,days agrees with the \citet{becker2015wasp47K2} results, but again we significantly improve the precision on this measurement by a factor of 3.6.

\subsection{Radial Velocity Analysis}\label{sec:rv_analysis}
We modelled our ESPRESSO radial velocity data along with the archival data from HARPS-N, HIRES, and CORALIE using the \textsc{exoplanet} Python package \citep{exoplanet:exoplanet}. The \textsc{exoplanet} package allows for robust probabilistic modelling of astronomical time series data using \textsc{PyMC3}. We used \textsc{exoplanet} to model the multiple radial velocity data sets using the No U-Turns Sampler Hamiltonian Monte Carlo method.

The free system parameters included in the analysis were the orbital periods of the planets, $P_i$, the times of conjunction, $T_{C,i}$, and the radial velocity semi-amplitudes, $K_{i}$, where $i$ represents the planets $b,c,d, \&\, e$. For $P_i$ and $T_{C, i}$ we used Gaussian priors taken from the posteriors of the transit analysis.  We also fitted for the eccentricities and arguments of periastron of planets c and d, $e_c, e_d, \omega_c, \omega_d$. As with the transit analysis, we fix $e_b = e_e = 0$, and for $e_d$ we imposed a half-Gaussian prior with a width of 0.014 and centered on 0. For $e_c$ we use a uniform prior constraining the eccentricity to be between 0 and 1. For $\omega_c$ and $\omega_d$ we used uniform priors between -180$^{\circ}$ and +180$^{\circ}$. 
We also include a white noise jitter term, $\sigma$, and a systemic radial velocity, $\gamma$, for each instrument. For CORALIE, we use independent jitter and systemic velocity terms for the data taken before and after the upgrade in November 2014.

The archival PFS data \citep{dai2015wasp47PFS} has been excluded from prior analyses on \Nstar\ due to the presence of large scatter in the radial velocities and the high risk of contamination from systematic errors \citep{vanderburg2017wasp47}. From our initial modelling, we also find a large jitter term is required for the PFS data. Motivated by this, we investigated the effect had by excluding the PFS data from our analysis. We find that the derived parameter values, specifically $K_{i}$, are unaffected by including the PFS data, and that the maximum log-likelihood value obtained during the sampling increases when the PFS data are not included in the analysis. Therefore, we also exclude the PFS data from our analysis.

From this initial modelling, we found a significant jitter term was required for our ESPRESSO data. The value required was $\sigma_{\rm ESP} = 3.25$\,ms$^{-1}$, compared to the median photon-limited uncertainty of 0.55\,ms$^{-1}$ for the ESPRESSO radial-velocities. Motivated by this we searched for evidence of periodicity in the residuals to the initial model. This excess scatter in the ESPRESSO radial-velocities suggests the presence of additional noise that we have not accounted for in our model. This additional noise is likely to arise from the stellar activity of \Nstar. We studied the available data for \Nstar\ to investigate any evidence for stellar variability.

\subsection{Stellar Rotation Analysis}\label{sec:rot_analysis}
\citet{vanderburg2017wasp47} used the HARPS-N spectra to derive a maximum stellar rotational velocity of 2\,kms$^{-1}$ based on the line broadening of the stellar absorption lines.  This implies a minimum rotation period of \protmin\ = \Nprotmin\,days. Independently, \citet{sanchisojeda2015wasp47RM} derived a value of $v\sin i = 1.8^{+0.24}_{-0.16}$\,km\,s$^{-1}$ from their Rossiter-McLaughlin analysis of \Nplanetb. From this measurement we estimate a rotation period of \Nstar\ of \prot\ = 31.96$\pm$4.36\,days. These \prot\ estimates are assuming \Nstar\ is aligned along our line--of--sight. This is a reasonable assumption given the Rossiter-McLaughlin measurement of a spin aligned orbit for \Nplanetb\ \citep{sanchisojeda2015wasp47RM}. These \prot\ limits and estimates will be important for properly interpreting our rotation analyses.

\subsubsection{Photometric Rotation Analysis}\label{sec:phot_rot_analysis}
\begin{figure}
    \centering
    \includegraphics[width=\columnwidth]{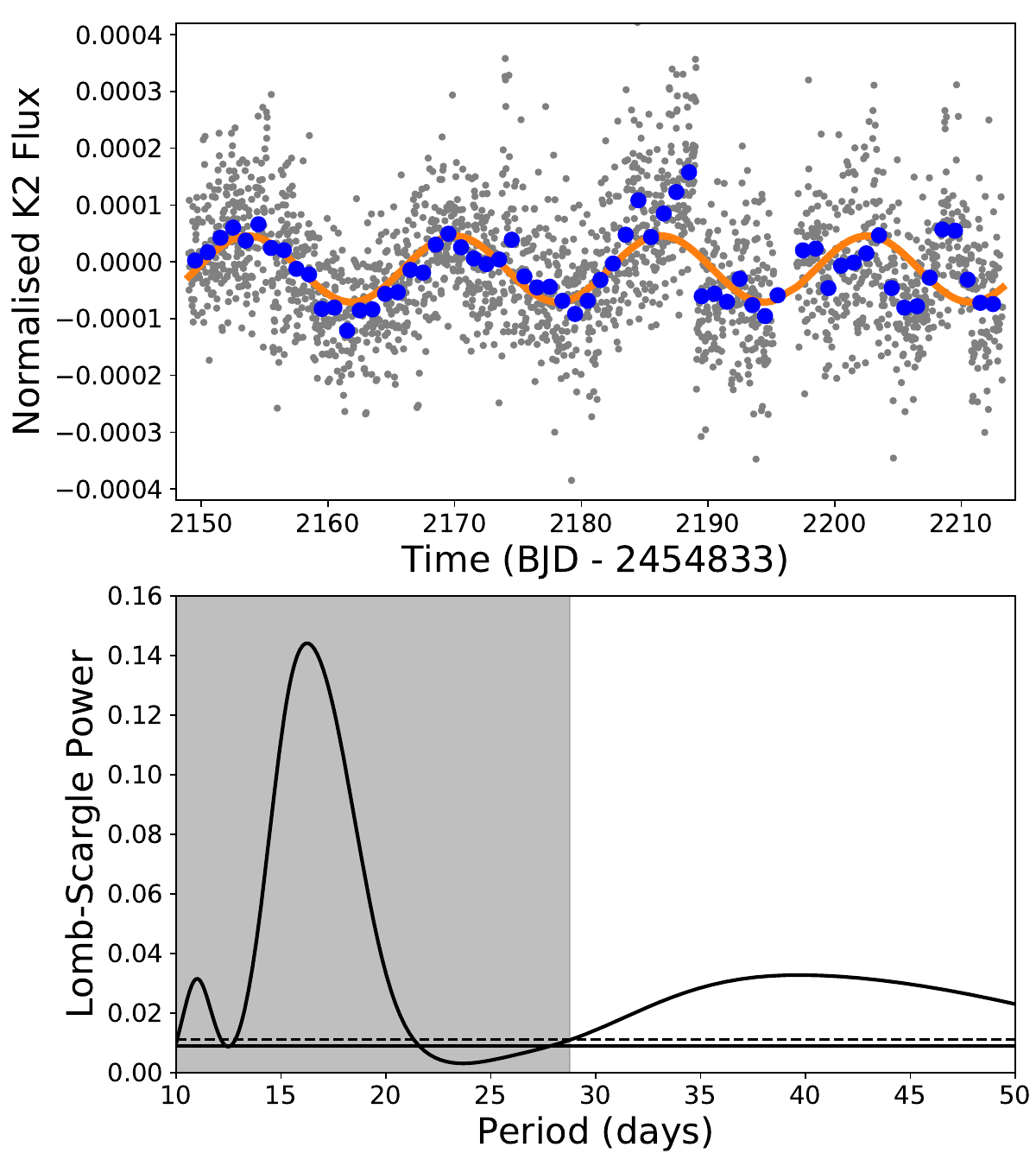}
    \caption{\textbf{Top:} K2 photometry from campaign 3 for \Nstar\ used for the rotation analysis in this work. The grey points show the detrended data (see Section~\ref{sec:phot_rot_analysis}) and the blue points give the data binned to a timescale of 1\,day. The orange line shows the most significant Lomb-Scargle period of 16.26\,days with an amplitude of 0.1\,mmag. \textbf{Bottom:} Lomb-Scargle periodogram for the K2 photometry. As with Fig.~\ref{fig:pgram} the shaded area highlights the rotation periods excluded by the minimum \prot\,=\protmin. The significant peak at 16.26\, days likely corresponds to half the true rotation period. The horizontal lines give the 0.1\% (solid) and 0.01\% (dashed) false alaram probabilities respectively.}
    \label{fig:k2_phot}
\end{figure}
We used the K2 data of \Nstar\ to search for photometric signs of stellar activity. Stellar spots are cooler and less bright than the majority of the stellar surface. As the star rotates, these spots rotate into and out of view of the telescope, resulting in a sinusoidal-like variation in the photometric data at the rotation period of the star, or some harmonic of this period. During the production of the K2 short cadence light curve, long term photometric variations were fitted and removed using a spline \citep{becker2015wasp47K2}. It is these exact variations that we want to study here. Therefore, we use a different long cadence K2 light curve that had been reduced using the "self-flat-fielding" method \citep{vanderburg2014k2sff}. This reduction technique significantly improves upon the precision of the raw K2 photometry and in most cases gets to a precision within a factor of two of the photometry delivered by Kepler during the nominal mission \citep{vanderburg2014k2sff}. More importantly, this method preserves long term astrophysical variations in the photometry.

We used the orbital parameters from \citet{becker2015wasp47K2} to remove any data points taken during a transit of any of the three \Nstar\ transiting planets. We also excluded the data points during two sharp ramps. These two ramps are related to the settling of the roll of the spacecraft at the start of the campaign and after the change in the roll direction around 50\,days into the campaign.   We fit quadratic polynomials to the two remaining continuous data chunks in order to remove long-term systematic trends in the data believed to be spacecraft systematics rather than astrophysical in nature. The detrended K2 photometry is shown in Figure~\ref{fig:k2_phot}.

We ran a Lomb-Scargle analysis on the remaining out-of-transit photometry. The resultant periodogram is shown in the bottom panel of Figure~\ref{fig:k2_phot}. The K2 photometry displays some sign of periodicity on periods which are similar to those present in the ESPRESSO data. However due to the 67\,day length of the K2 photometry and the detrending methods applied, this period search becomes less sensitive for longer period signals, especially for signals around half the monitoring length and longer. The significant peak in the K2 periodogram is at a period of 16.26$\pm$1.94\,days. This period is shorter than the \protmin\ limit set by \citet{vanderburg2017wasp47}. Periodic signals are expected in photometric time series at the second harmonic of the rotation period \citep{clarke2003photvariability}. Particularly in cases where there are multiple active regions of the surface of the star where the largest peak in the periodogram can be at half the true rotation period \citep[eg. ][]{mcquillan2013stellarrot}. Therefore, this 16.26$\pm$1.94
\,day signal likely corresponds to half the true rotation period of \Nstar, giving a rotation period of \prot\,=\,32.5$\pm$3.9\,days, which is consistent with the prediction from the \citet{sanchisojeda2015wasp47RM} $v\sin i$ measurement.

We also turn to theoretical predictions for stellar rotation periods to confirm that this value is a physically reasonable rotation period for \Nstar. We use the model from \citet{barnes2007fieldstargyrochron} which estimates \prot\ given the stellar magnitudes in the $B$ and $V$ pass-bands and the age of the star. For \Nstar, we have magnitudes of $V =$\,11.936\,mag and $B =$\,12.736\,mag, and by assuming a solar age of 4.5\,Gyr we find a predicted rotation period of \prot$_{\rm ; \ pred} \, = \, 34.96$\,days.

We note that \citet{hellier2012wasp47} searched the WASP photometry of \Nstar\ for signs of stellar rotation and did not find any significant rotational modulation, placing an upper limit of 0.7\,mmag for the amplitude of any such modulation. From the K2 photometry in Figure~\ref{fig:k2_phot}, the peak-to-peak amplitude of the modulation is approximately 0.1\,mmag, therefore the non-detection of rotational modulation in the WASP photometry is not inconsistent with the K2 detection. Similarly, due to the short time coverage and low precision, the TESS data, is unable to help constrain the stellar rotation.

\subsubsection{Spectroscopic Rotation Analysis}\label{sec:spec_rot_analysis}
\begin{figure}
    \centering
    \includegraphics[width=\columnwidth]{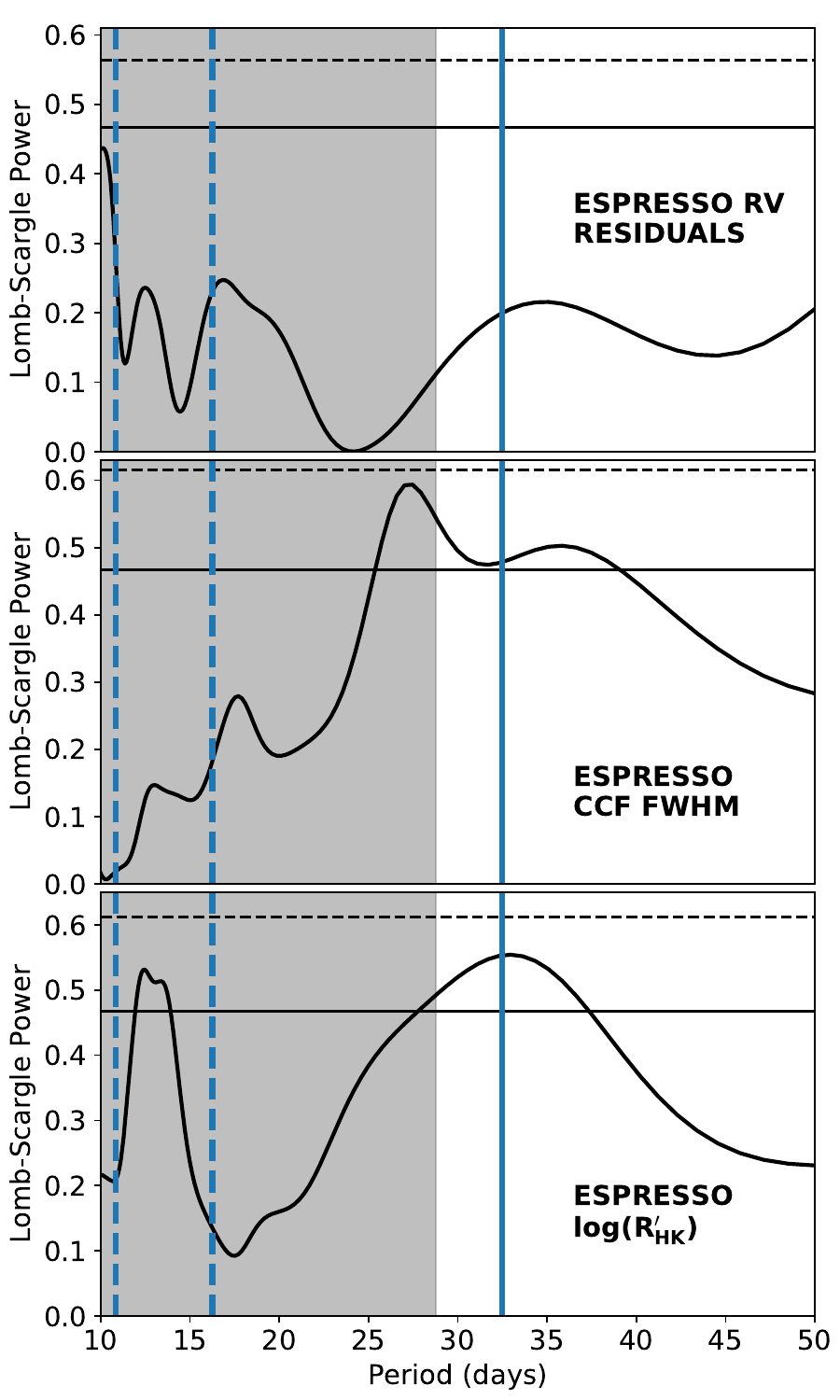}
    \caption{Generalized Lomb-Scargle periodograms for: \textbf{Top:} ESPRESSO radial velocity residuals to the initial model from Section~\ref{sec:rv_analysis}; \textbf{Middle:} the ESPRESSO CCF FWHM measurements; \textbf{Bottom:} measurements of \logrhk\ from the ESPRESSO spectra. The solid vertical blue line denotes the rotation period of \prot\ = 32.5\,days derived from the K2 photometry, with the vertical dashed blue lines marking \prot/2 and \prot/3 harmonics. The shaded area highlights stellar rotation periods that are excluded by the calculated minimum rotation period of \Nprotmin\,days. The horizontal lines indicate the 1\% (dashed) and 10\% (solid) false alarm probabilities.}
    \label{fig:pgram}
\end{figure}
\begin{figure}
    \centering
    \includegraphics[width=\columnwidth]{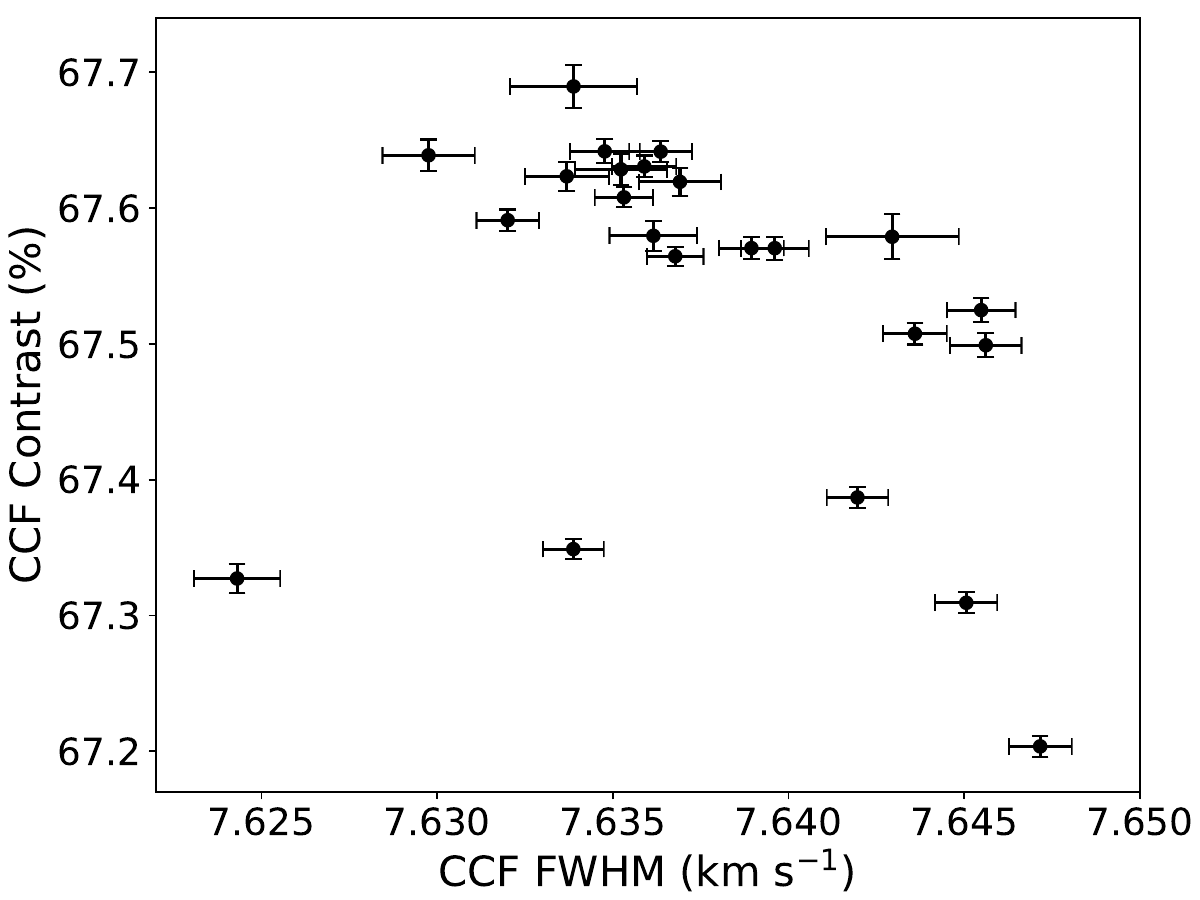}
    \caption{The contrast and FWHM values for the ESPRESSO cross-corelation functions for our observations of \Nstar.   We see a strong anti-correlation, indicative of spot-induced stellar activity \citep{boisse2011stellaractivity}.}
    \label{fig:fwhm_vs_cont}
\end{figure}

We search for periodic signals in our spectroscopic ESPRESSO data that might provide evidence that the excess scatter seen in the ESPRESSO radial velocities arises as a result of stellar activity. We ran a generalized Lomb-Scargle analysis on the residuals to the model from the fitting in Section~\ref{sec:rv_analysis}, which revealed periodicities at periods of roughly 35 and 18 days, as shown in Figure~\ref{fig:pgram}. We also searched for periodic signals in the ESPRESSO CCF FWHM measurements, as the FWHM of the CCF has been shown to act as a stellar activity indicator \citep{boisse2011stellaractivity, oshagh2017activityindicators}. Stellar spots suppress the flux contributions from different sections of the stellar surface to the stellar lines and thus and modify the CCF profile. Spots on the limb of the star suppress the wings of the line, resulting in the CCF profile showing a smaller FWHM.  Conversely, spots on the center of the stellar disc affect the center of the stellar lines, resulting in a wider CCF FWHM \citep{boisse2011stellaractivity}. Periodic signals with the same periods as those found in the RV residuals were found in the CCF FWHM measurements (see Figure~\ref{fig:pgram}). \citet{boisse2011stellaractivity} also find that the stellar activity causes the CCF contrast, the fractional height of the CCF peak, to anti-correlate with the FWHM. We compare the contrast and FWHM for the ESPRESSO CCFs and find a strong anti-correlation (see Figure~\ref{fig:fwhm_vs_cont}). This anti-correlation is in line with the predictions and strengthens the confidence that the variations in the CCF FWHM, and by extension the ESPRESSO radial velocity residuals, are a result of stellar activity.

We also used the Ca-II H and K lines in the ESPRESSO spectra to determine values of the activity index \logrhk. Running a further Lomb-Scargle analysis on this activity indicator reveals a periodic signal in the range 30-40\,days with a peak around 33\,days (Figure~\ref{fig:pgram}).

\subsection{Stellar Activity Analysis}\label{sec:gp_analysis}
We expect the spectroscopic signals from stellar activity to manifest at periods equal to \prot\ and at the \prot/2 and \prot/3 harmonics \citep{boisse2011stellaractivity}.  From the ESPRESSO radial velocity periodogram (top panel of Figure~\ref{fig:pgram}), we see a signal close to P=35\,days, with harmonics close to P/2 and P/3.  These peaks are also seen in the CCF FWHM periodogram, although an extra peak is seen close to 27\,days that is likely due to the moon. The \logrhk\ activity index from the ESPRESSO spectra has a peak in the periodogram at approximately 33\,days.

The periodicities detected in the ESPRESSO RV residuals and activity indicators are consistent with the 32.5$\pm$3.9\,day rotation period derived from the K2 photometry for \Nstar. Based on the continuous coverage and very high precision of the K2 data, we take the K2 \prot\ value as the most probable rotation period.
\begin{figure*}
    \centering
    \includegraphics[width=\textwidth]{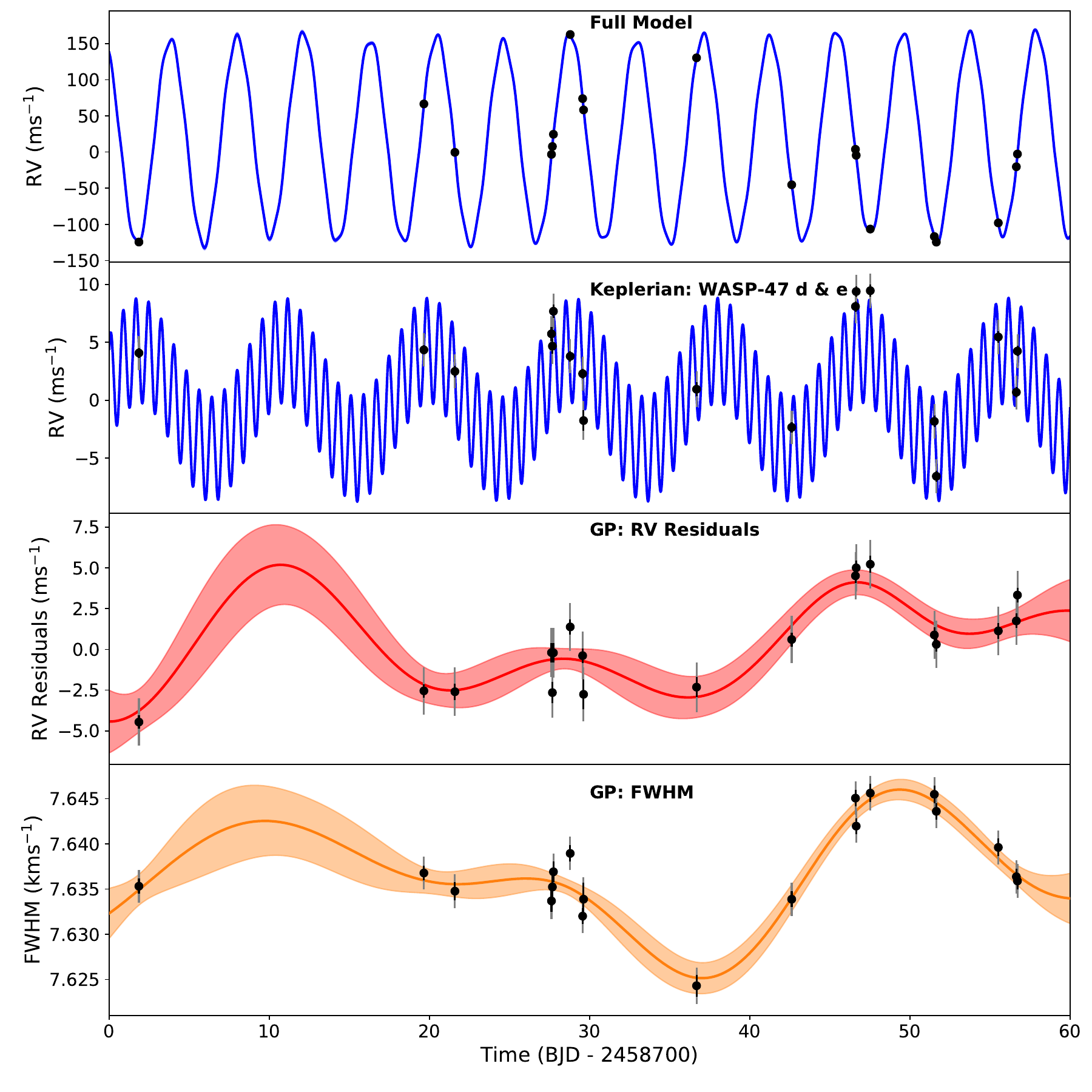}
    \caption{ESPRESSO time series data for \Nstar. \textbf{Top:} ESPRESSO radial velocity measurements with the systemic velocity $\gamma_{\rm ESP}$ subtracted. The blue line gives the full Keplerian model for all four planets and the GP. \textbf{Second:} ESPRESSO radial velocity measurements with $\gamma_{\rm ESP}$, the Keplerian models for \Nplanetb\ and c, and the GP model subtracted. The blue line gives the two-planet Keplerian model for \Nplanetd\ and e. \textbf{Third:} ESPRESSO radial velocity residuals to the four-planet Keplerian model. The red line gives the GP model and the shaded area provides the 1$\sigma$ confidence interval of the GP. \textbf{Bottom:} FWHM measurements of the ESPRESSO CCFs. The orange line gives the GP model and the shaded area provides the 1$\sigma$ confidence interval of the GP.}
    \label{fig:rv_esp}
\end{figure*}

\subsection{Gaussian Process Analysis}\label{sec:gp_analysis}
We utilize a Gaussian Process (GP) kernel with a periodicity close to \prot\,=\,32.5\,days in order to accurately model the radial velocity variability due to spot rotation. We use the {\sc celerite2} {\it rotation term} kernel constructed from two Simple Harmonic Oscillator (SHO) terms at \prot\ and \prot/2, implemented using the {\sc exoplanet} \citep{exoplanet:exoplanet} and {\sc celerite2} \citep{celerite1, celerite2} Python packages. A single SHO term is given by
{\begin{equation}
    \small
    \kappa(\tau) = \sigma_{GP}^2 \exp\left(-\frac{\pi\tau}{Q P_{\rm Rot}}\right) \left[\cos\left(2\pi\eta\frac{\tau}{P_{\rm Rot}}\right) + \frac{1}{2\eta\,Q}\sin\left(2\pi\eta\frac{\tau}{P_{\rm Rot}}\right)\right]
\end{equation}}
where $\eta = |1-(4Q^2)^{-1}|^{1/2}$. The {\it rotation term} kernel takes as its hyperparameters: the standard deviation of the process, $\sigma_{GP}$, which is related to the amplitude of the variability signal, the stellar rotation period, \prot, the signal quality, $Q$, the difference in signal quality between the \prot\ and \prot/2 modes, $\Delta Q$, such that $Q_{P_{\rm Rot}} = Q_{P_{\rm Rot}/2} + \Delta Q$, and the mix factor between the two modes, $f$, such that $\sigma_{P_{\rm Rot}}^2 = f\,\sigma_{P_{\rm Rot}/2}^2$. Such a kernel has been designed to be a good model for variability due to stellar rotation and has been used successfully in previous exoplanet high precision radial velocity analysis \citep[eg.][]{osborn2021GPtoi755}.

During the analysis, we used the same kernel to fit the radial velocity measurements and the CCF FWHM activity indicators of the ESPRESSO observations. By modelling the variations in CCF FWHM simultaneously with the RVs, we get a better estimate of the impact of the stellar variability on the RV measurements. This allows us to limit any impact on the measured planetary RV signals due to over-fitting of the GP, and so extract better quality mass measurements. A similar method was used by \citet{osborn2021GPtoi755} to account for the stellar variability of TOI-755.

For this method, during the sampling we use the same \prot, $Q$, $\Delta Q$, and $f$ hyperparameters for all data sets and a different $\sigma_{GP}$ and mean for each activity or radial velocity time series. We use a wide Gaussian prior centered on 35\,days for \prot\ taken from our Lomb-Scargle analysis. The $Q$ hyperparameter is related to the damping timescale of the SHO modes, which in a physical sense is related to the decay timescale of the stellar spots. As such, we also implement a prior enforcing $Q > \pi$, which in turn ensures the spot decay timescale is longer than \prot. This requirement comes from our prior knowledge of the lifetimes of active stellar regions \citep{donahue1997}. Avoiding low values of $Q$ also helps prevent the GP over-fitting the data \citep{kosiarek&crossfield2020}.

To assess the statistical justification of using a GP to account for stellar activity in the radial velocity data sets, we perform a simple model comparison analysis. We calculate and compare the Bayesian Information Criterion \citep[BIC, ][]{schwarz1978bic, neath2012bic} for both sets of models, one with the GP and one without. This statistic assesses whether or not the model fit to the data is sufficiently improved to justify the increased complexity of the new model. When comparing models, the model which produces the lower BIC value is preferred.

We calculate independent BIC values for the ESPRESSO, HARPS-N, and HIRES datasets. We calculate the change in the BIC, $\Delta\,{\rm BIC}$, between the models with and without the GP included. We find values for the HARPS-N and HIRES data sets of $\Delta\,{\rm BIC}_{\rm HARPS-N} = 4.88$ and $\Delta\,{\rm BIC}_{\rm HIRES} = 8.76$. This indicates that the inclusion of GPs for these two data sets is not justified. For the ESPRESSO radial velocities, we calculate a value of $\Delta\,{\rm BIC}_{\rm ESPRESSO} = -16.86$. This is strong statistical evidence for the inclusion of a GP to model the stellar activity in the ESPRESSO radial velocities. It is not unexpected that the use of a GP is justified for ESPRESSO and not HARPS-N and HIRES. The larger telescope diameter of the VLT results in the photon-limited uncertainties of the ESPRESSO radial velocities (0.55 ms$^{-1}$) being a factor of six smaller than those for HARPS-N (3.2 ms$^{-1}$) and a factor four smaller than HIRES (2.0 ms$^{-1}$). This increased precision of the data allows for the robust detection of the stellar activity signal. We do not use GPs for the CORALIE data as uncertainty is 12.5\,ms$^{-1}$, which is too large to detect the stellar activity signal seen in the ESPRESSO data.

\subsection{Final Combined Model}\label{sec:final_analysis}
We modelled the ESPRESSO, HARPS-N, HIRES, and CORALIE data using our final model built from Keplerian orbits for the four planets along with a GP to account for the stellar variability signals present in the ESPRESSO data set. The GP was simultaneously fit to the ESPRESSO radial velocities and CCF FWHM measurements, in order to limit the effect of over-fitting the GP to the radial velocities. We perform the sampling using the method detailed in Section~\ref{sec:rv_analysis}. We ran 40 chains for 10000 steps each, following a burn-in of 4000 steps for each chain. We calculate the Gelman-Rubin statistic \citep[\^R;][]{gelmanrubin92} for all the chains, and find that all chains have \^R $\lesssim\,1.001$ indicating good convergence.

From this analysis we find a radial velocity semi-amplitude for \Nplanete\ of $K_e =$ \NKe\,ms$^{-1}$. This value is consistent with the semi-amplitude derived from the initial modelling in Section~\ref{sec:rv_analysis} of $4.73 \pm 0.39$\,ms$^{-1}$, but represents an improvement in the precision of the measurement. This corresponds to an improvement in the mass measurement precision from \mpl\,=\,$7.03 \pm 0.59$\,\mearth\ for the initial model to \mpl\,=\,\Nmasse\,\mearth\ for the final model. We also note that the value of $P_b$ derived from this analysis is consistent with the value derived from the transit analysis in Section~\ref{sec:trans_analysis}.  The stellar rotation period of \Nstar\ derived from this analysis is \prot\,=\,\Nprot\,days, which again is consistent with the rotation period derived from the K2 photometry.

\begin{figure}
    \centering
    \includegraphics[width=\columnwidth]{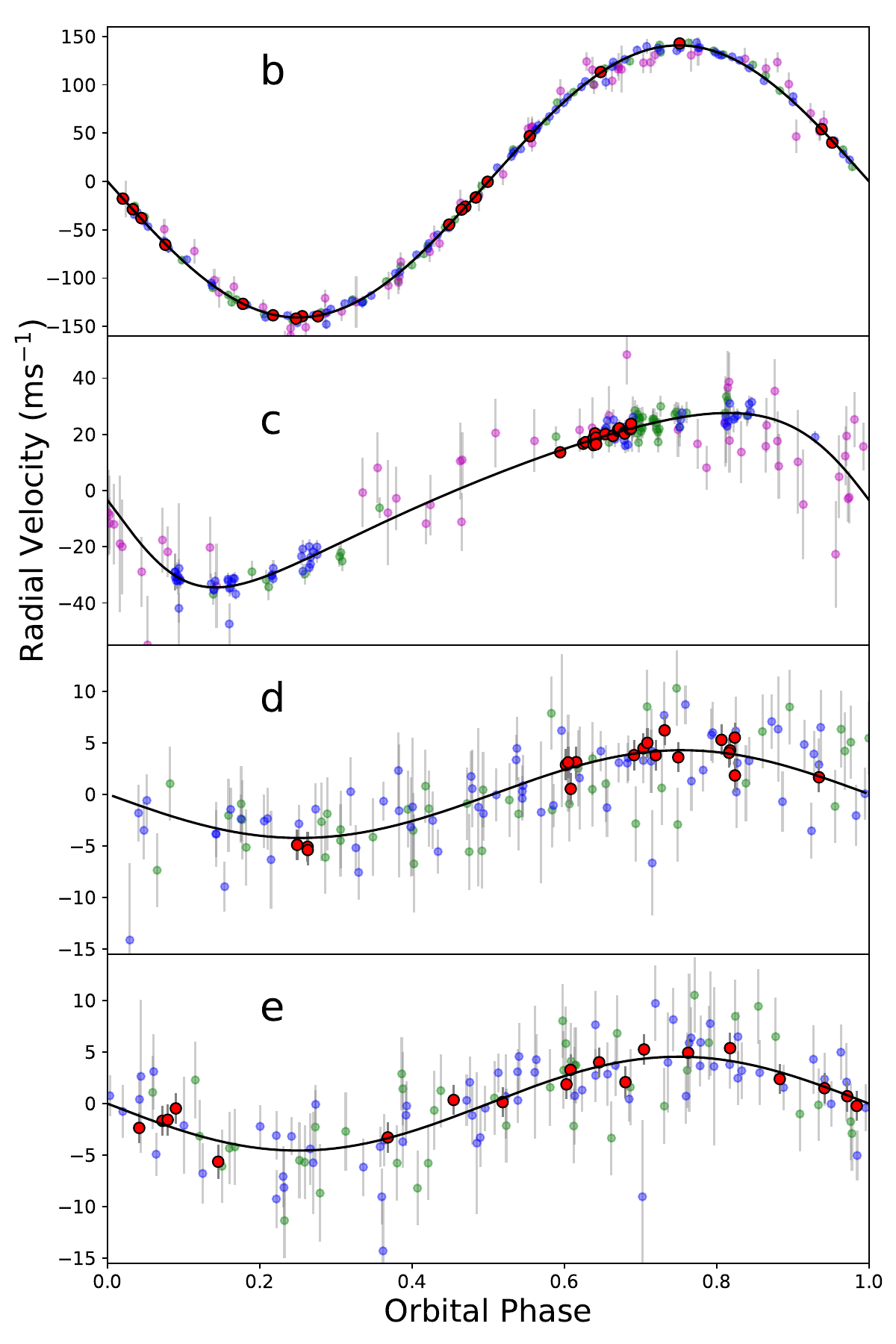}
    \caption{Phase-folded RVs with all GPs and systemic velocities removed for \Nplanetb\ (top), \Nplanetc\ (second), \Nplanetd\ (third), and \Nplanete\ (bottom). The colors of the data from the various instruments are: ESPRESSO (red); HARPS-N (blue); HIRES (green); CORALIE (magenta). Note we do not plot the CORALIE data for planets d and e as the RMS of the CORALIE data is larger than the semi-amplitudes of these two planets.}
    \label{fig:rv_phase}
\end{figure}

The ESPRESSO data are plotted in Figure~\ref{fig:rv_esp} and the radial velocity phase folds for all planets and instruments are plotted in Figure~\ref{fig:rv_phase}. The parameters derived are given in Tables~\ref{tab:params} \& \ref{tab:params2}.

With the inclusion of our ESPRESSO data we have obtained a mass of \Nplanete\ of \Nmasse\,\mearth. This value is consistent at the 1$\sigma$ level with the mass determined by \citet{vanderburg2017wasp47}, but we have improved on the precision they quote on the mass by 15\%. With the improved precision of the ESPRESSO data we have uncovered a clear stellar activity signal, which is consistent with the stellar variability seen in the K2 photometry. This allows us to model the stellar activity using a physically motivated GP informed by our knowledge of the stellar rotation.

Our results also allow us to improve the constraints on the bulk density of \Nplanete. We derive a density of $\rho_e = $\,\Ndensitye\,\gccc, making \Nplanete\ the super-Earth with the second most precisely constrained density to date, behind only \cnce\ \citep{bourrier201855cnce}. Our ESPRESSO data also improves the constraint on $M_d$. We find a mass of \Nmassd\,\mearth, which is both consistent with the mass measured by \citet{vanderburg2017wasp47} and a 13\% improvement on the precision of the mass measurement. For the \hj\ \Nplanetb\ we improve the precision of the radial velocity semi-amplitude, yet the precision on the mass remains unchanged from the \citet{vanderburg2017wasp47} measurements. From this, we conclude that the primary limiting factor for improving the constraints on $M_b$ arise from the constraints on the stellar mass, which is currently measured to 3\% precision, compared to the 0.3\% precision to which we have measured $K_b$. We do not achieve significant improvement on the measurement of the minimum mass of \Nplanetc. This is unsurprising, as our ESPRESSO radial velocity measurements only cover a small fraction of the orbital period of of \Nplanetc\ ($P_c =$ \Nperiodshortc\,days).

\subsection{Transit Timing Analysis}\label{sec:ttv_analysis}
\begin{figure*}
    \centering
    \includegraphics[width=0.95\textwidth]{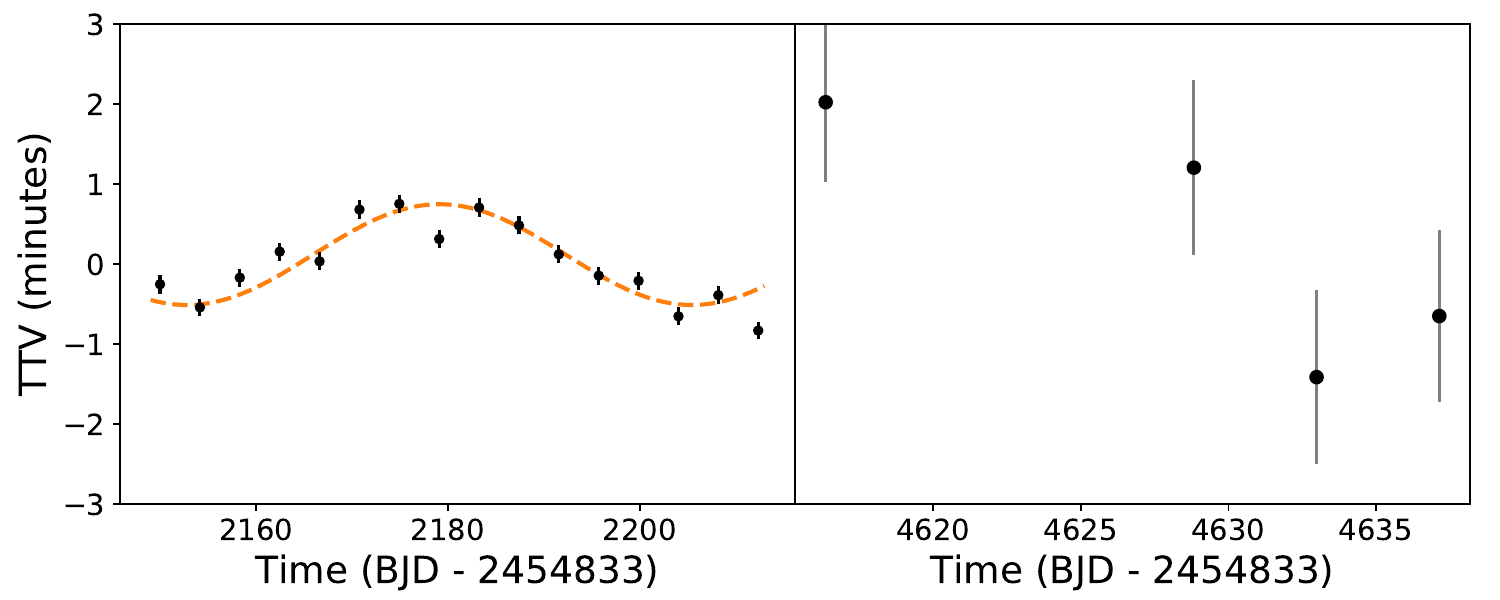}
    \caption{Observed TTVs for \Nplanetb\ for the transits in the \textbf{Left:} K2 photometry and \textbf{Right:} TESS photometry. The orange line gives the TTV model from \citet{becker2015wasp47K2}.
    }
    \label{fig:w47b_ttvs}
\end{figure*}
\Nplanetb\ and \Nplanetd\ have already been shown to exhibit significant transit timing variations \citep[TTVs, eg.][]{becker2015wasp47K2, weiss2017wasp47rvttv}. Due to the low signal-to-noise of the TESS transits of these planets, we are unable to strongly constrain the transit center times for the individual TESS transits of \Nplanetd\ or \Nplanete.  However we can measure the transit times for the hot Jupiter \Nplanetb.

\begin{table}
    \centering
    \begin{tabular}{cc}
    \noalign{\smallskip} \noalign{\smallskip} \hline  \hline \noalign{\smallskip}  
    Epoch & Mid-Transit Time \\
          &  (BJD TDB) \\
    \hline 
     587 & 2459449.35411 $\pm$ 0.00069 \\
     590 & 2459461.83099 $\pm$ 0.00076 \\
     591 & 2459465.98832 $\pm$ 0.00076 \\
     592 & 2459470.14800 $\pm$ 0.00075 \\
    \hline  
    \end{tabular}
    \caption{Measured TESS mid-transit times for \Nplanetb. The epoch is relative to the $T_C$ value given in Table~\ref{tab:params}.}
    \label{tab:tess_tcs}
\end{table}

We re-fit each transit of \Nplanetb\ in the TESS data in order to measure the individual $T_C$ values. During this analysis, we fix the transit shape to the model derived in Section~\ref{sec:final_analysis}. The same model is used to subtract the transits of \Nplanetd\ and \Nplanete\ from the light curve before fitting. We also perform the same analysis for the \Nplanetb\ transits in the short cadence K2 data. The transit times measured are plotted in Figure~\ref{fig:w47b_ttvs} and provided in Table~\ref{tab:tess_tcs}. The uncertainty on an individual $T_C$ from the TESS data is on the order of 1\,minute, which is roughly twice the TTV amplitude seen in the K2 data. The TESS transit times display a scatter of a similar magnitude. As such, the $T_C$ measurements are consistent with the TTV predictions from the K2 results (see Figure~\ref{fig:w47b_ttvs}).  However the detection is marginal, and we cannot know the phase of the signal due to the significant time gap between K2 and TESS data.

\section{Discussions}
\begin{figure}
    \centering
    \includegraphics[width=\columnwidth]{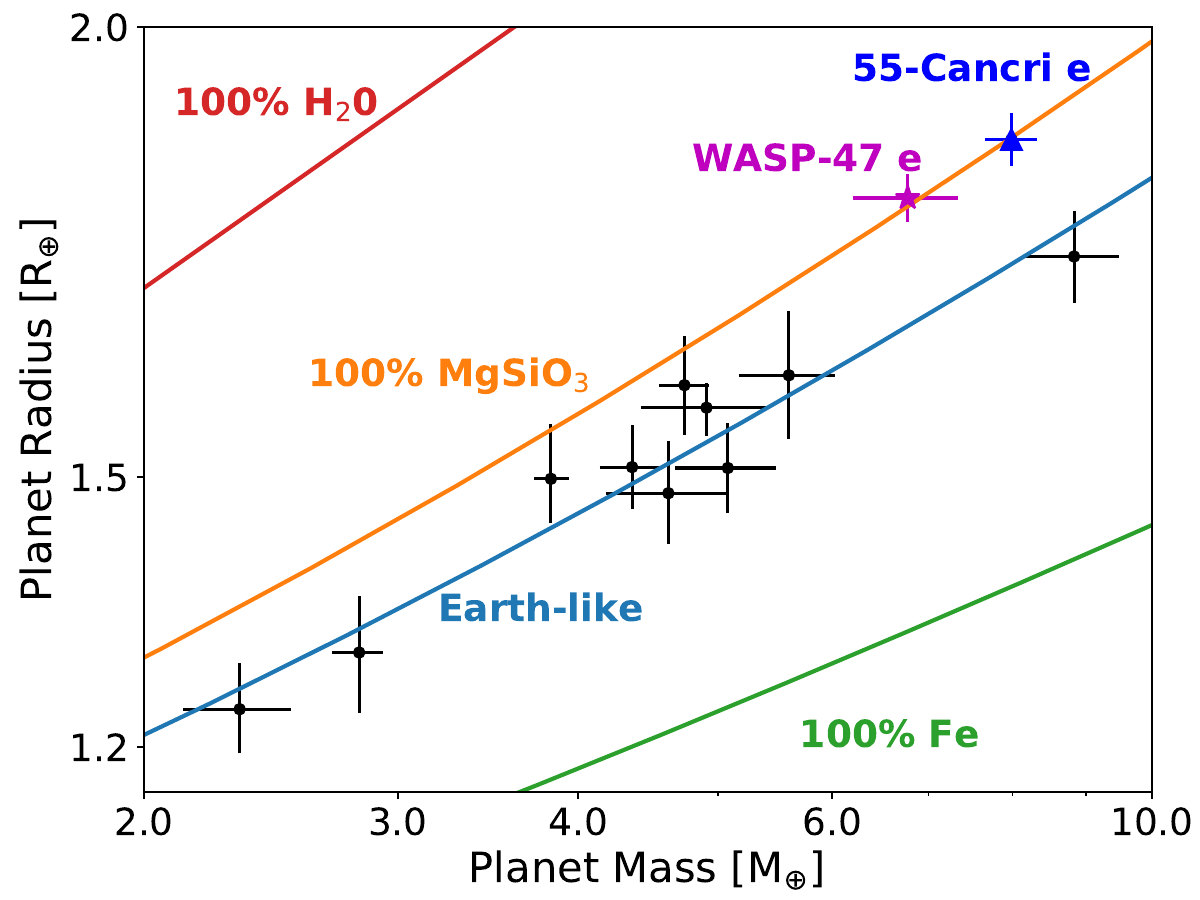}
    \caption{Mass-Radius diagram for low-mass exoplanets. The measurements and uncertainties are taken from the NASA exoplanet archive (accessed 2021 May 12). The black points show exoplanets with masses and radii measured to better than 10\% precision. \Nplanete\ is shown with the magenta star, and \cnce\ is plotted with the blue triangle. These two planets now have the most precise density measurements for any known super-Earths.  The solid lines give various composition models from \citet{zeng2016mrmodels, zeng2019mrmodels}.}
    \label{fig:MR}
\end{figure}

\begin{figure}
    \centering
    \includegraphics[width=\columnwidth]{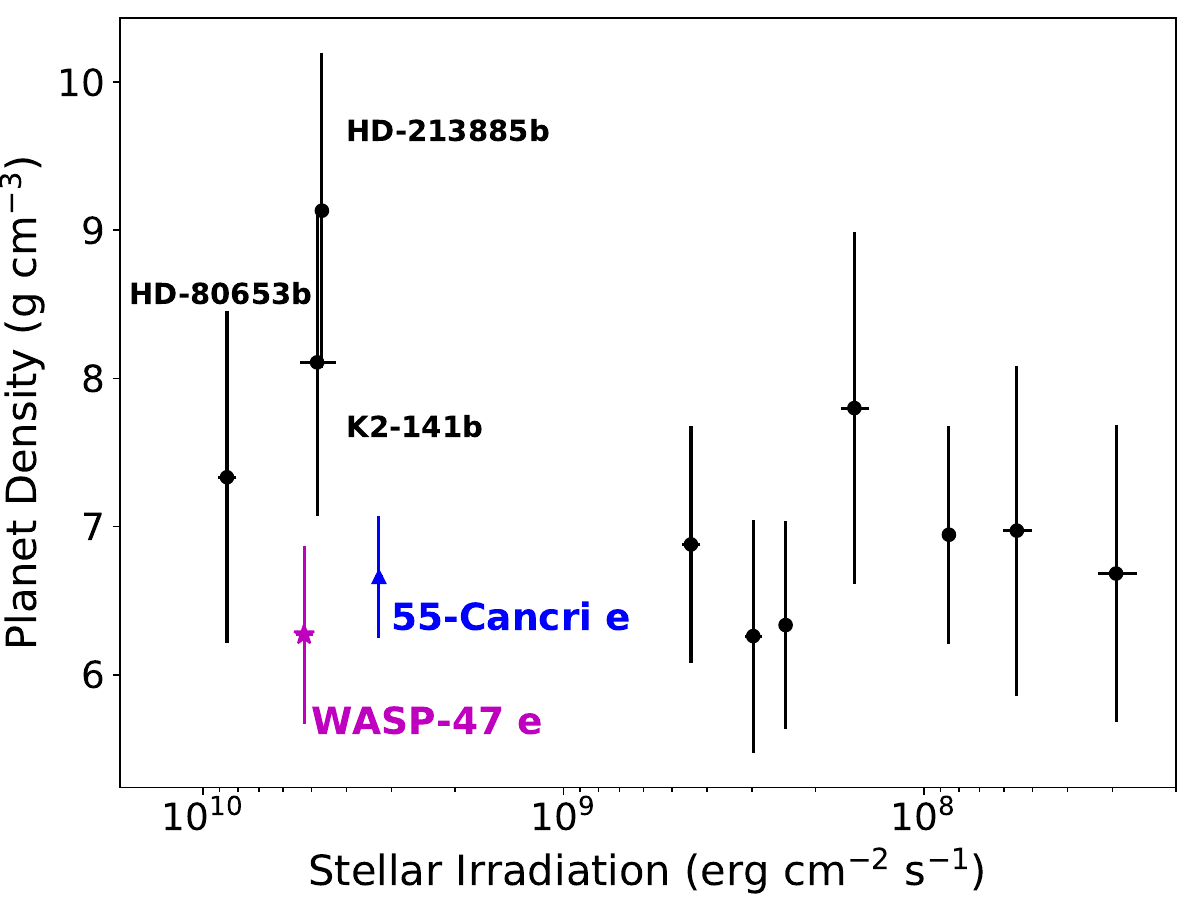}
    \caption{The variation of bulk planetary density with stellar irradiation for the sample of planets plotted in Figure~\ref{fig:MR}. \Nplanete\ is depicted by the magenta star and \cnce\ by the blue triangle.}
    \label{fig:irrad_mass_den}
\end{figure}

The mass and density we measure for \Nplanete\ are consistent at the 1$\sigma$ level with the results of \citet{vanderburg2017wasp47}. \citet{vanderburg2017wasp47} use the planetary composition models of \citet{lopez2017uspmodels} to show that their measured parameters of \mpl\,= $6.83 \pm 0.66$\,\mearth\ and \rhopl\,= $6.35 \pm 0.64$\,\gccc\ are consistent with \Nplanete\ having a steam-rich layer surrounding an Earth-like core and mantle. Due to the agreement between these measurements the the values derived in this work, such a composition remains a plausible scenario.

We compare our newly derived parameters for \Nplanete\ to the sample of known planets with \mpl\,$< 10$\,\mearth, \rpl\,$< 2$\,\rearth\ and mass and radius measurement precisions better than 10\% (see Figure~\ref{fig:MR}). We find that \Nplanete\ is similar in mass and radius to \cnce, and that these two exoplanets are the only in the sample which onto the 100\% MgSiO$_3$ composition model of \citet{zeng2019mrmodels}. The \Nstar\ and \cnc\ planetary systems are also the only two to contain close in giant planets: \Nplanetb\ (\mpl\ = \Nmassbjup\,\mjup; P = \Nperiodshortb\,days) and \cnc\ b (\mpl\ = 0.804$\pm$0.009\,\mjup; P = 14.6516\,days). \Nplanete\ has a lower density than K2-141\,b and HD-213885\,b despite receiving a very similar amount of stellar irradiation (see Figure~\ref{fig:irrad_mass_den}). \cnce\ receives a similar level of irradiation and has a similar density to \Nplanete. It is possible that the presence of the close-in giant planet companions to these super-Earths has caused them to have a lower density than other planets with similar levels of stellar irradiation. 

Due to the intense stellar irradiation, this reduced density is almost certainly not due to an extended H/He atmosphere because such an atmosphere would have been lost through photo-evaporation \citep{penz2008exomassloss, sanzforcada2011photoevap}. One possibility could be the scenario with a steam-rich layer proposed by \citet{vanderburg2017wasp47}. Alternatively, \citet{dorn2019superearthformation} recently proposed that \Nplanete\ and \cnce\ could have compositions rich in refractory elements, such as Ca and Al, which condense out of protoplanetary disks at high temperatures. This different elemental make-up was shown to result in planets with densities 10-20\% less than a planet of the same \rpl\ but with an Earth-like composition \citep{dorn2019superearthformation} and therefore would explain the lower densities of these two planets, given the irradiation they receive.

\begin{figure}
    \centering
    \includegraphics[width=\columnwidth]{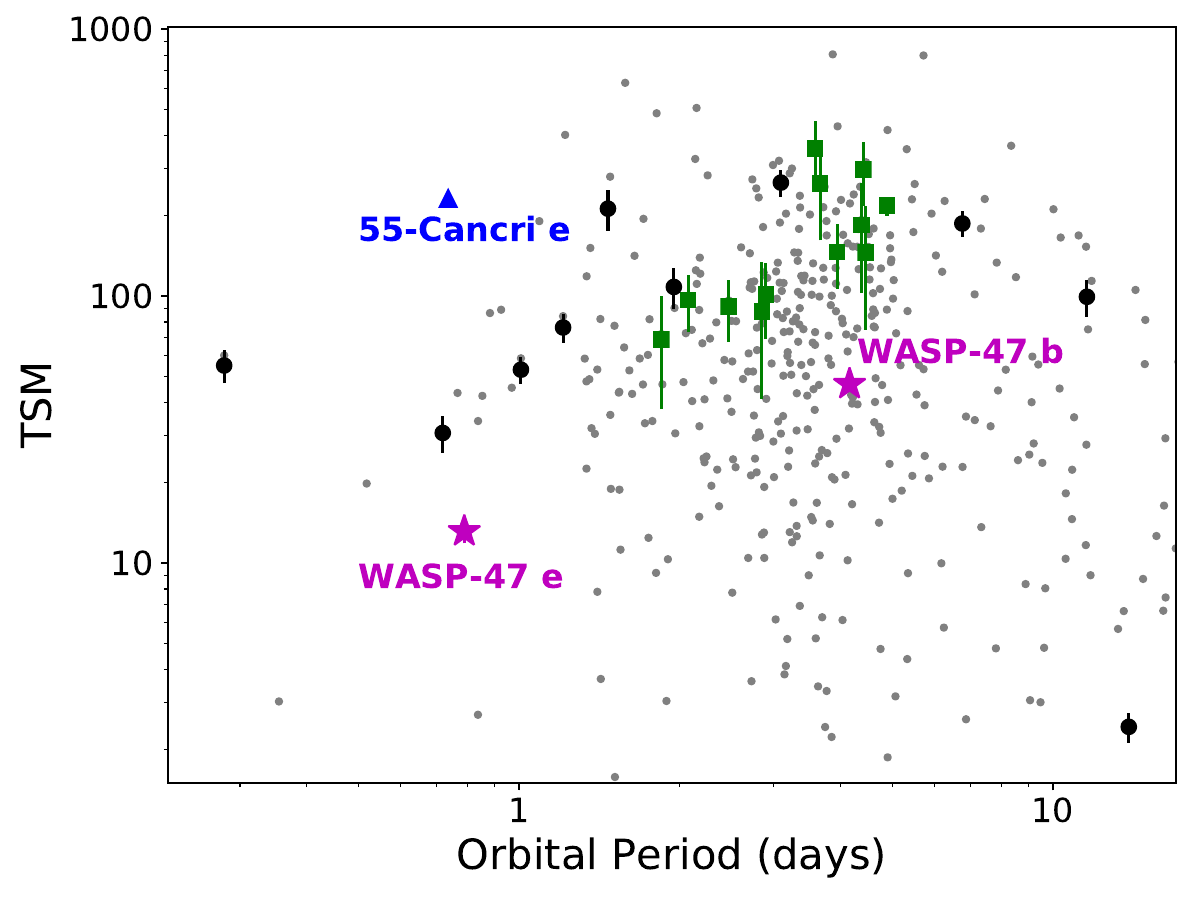}
    \caption{Comparison of TSM vs orbital period for various known exoplanets. We plot the sample of planets from the NASA Exoplanet Archive with a radius measured to better than 10\% precision and a mass measured to better than 50\% precision (grey points), the JWST community targets \citep[green squares; ][]{stevenson2016jwstcomm} and the sample of planets from Figure~\ref{fig:MR} (black points). \Nplanetb\ and \Nplanete\ are depicted by the magenta stars and \cnce\ by the blue triangle.}
    \label{fig:w47be_jwst}
\end{figure}
Obtaining transmission spectroscopy observations for the three inner \Nstar\ planets will provide further clues to unlock the formation history of this system. We use the transmission spectroscopy metric (TSM) from \citet{kempton2018tsm} to assess the potential for obtaining such observations. The TSM provides near-realistic values of the expected signal-to-noise ratio obtained from a 10\,hr observation sequence with JWST. For the inner planets, we calculate values of \Ntsme\ (\Nplanete), \Ntsmb\ (\Nplanetb), and \Ntsmd\ (\Nplanetd). While modest compared to other planets with similar orbital parameters due to the relative faintness of \Nstar\ (V = 11.94\,mag) (see Figure~\ref{fig:w47be_jwst}), these values indicate that a significant amount of information on the atmospheric compositions of these planets can be achieved through JWST observations. Combined with the precise constraints on the masses of these planets, these atmospheric observations would provide great insight into the composition and possible formation history of the \Nstar\ planets.

The potential detection of an atmosphere on \cnce\ has been reported from both HST \citep{tsiaras201655cancrihstatmos} and Spitzer \citep{angelo201755cancrispitzeratmos} observations. We also note that \cnce\ will be observed by JWST during its Cycle 1 observations\footnote{JWST GO Programs \href{https://www.stsci.edu/jwst/science-execution/program-information.html?id=1952}{1952} and \href{https://www.stsci.edu/jwst/science-execution/program-information.html?id=2084}{2084}.}. Due to the similarities between the two planets and their environments, any revelations about the atmospheric conditions of \cnce\ have implications for the likelihood of an atmosphere on \Nplanete.

\Nplanete\ and \cnce\ are the only two planets to fall solidly on the 100\% MgSiO$_3$ composition line in Figure~\ref{fig:MR}. This different composition of \Nplanete\ and \cnce\ are possibly the result of these systems forming through a different pathway to the other planets shown. The nature of \Nplanete\ orbiting interior to a \hj\ also strongly suggests a different planetary formation and evolution mechanism to the large majority of \hj\ systems \citep{huang2016lonelyHJs}. The formation of hot Earths and Neptunes interior to \hjs\ can arise as a result of portions of the protoplanetary disk being shepherded to the inner regions of the planetary system by a giant planet migrating through disk interactions \citep{foggnelson2005, foggnelson2007typeImigration}. This disk shepherding prior to the formation of \Nplanete\ could provide a high temperature environment amenable to the formation from high temperature condensates \citep{dorn2019superearthformation}.

\citet{poon2021insituformation} demonstrated that not only can an in-situ formation mechanism produce planetary systems containing \hjs\ and inner small planets but also that this formation mechanism does not reproduce the observed population of single \hjs. This again suggests that the \Nstar\ planetary system formed through a different mechanism to other \hj\ systems. This scenario was also suggested by \citet{huang2016lonelyHJs}, who note that the \Nstar\ system bears stronger resemblance to the population of \wj\ systems, which often have smaller planets interior to the \wj. It is possible that the \Nstar\ and \cnc\ systems formed through a mechanism similar to \wj\ systems, resulting in different bulk compositions for \Nplanete\ and \cnce. 

With only two super-Earths with close giant planet companions and precisely measured densities, we do not have a large enough sample size from which to draw significant conclusions. A larger sample of known planets interior to \hjs\ would allow us to better identify any trends in the properties of these inner companions. Therefore, further discoveries of small planets interior to hot and \wjs\ are needed to shed light on how these systems form, and how the compositions the small planets are sculpted by their formation. There is a possibility that other known \hjs\ have small planets orbiting interior to them. The majority of known transiting \hjs\ were discovered by ground based transit surveys, and so the discovery data did not have sufficient photometric precision to detect super-Earth planets. \Nplanetd\ and \Nplanete\ are only known thanks to the K2 data and of the 536 known exoplanets with $P < 10$\,days and \mpl\,>\,0.1\,\mjup\ only 80 have received such high precision monitoring with Kepler or K2\footnote{NASA exoplanet archive, accessed 2022-01-20.}. It is therefore possible that several of the \hj\ host stars that have not been monitored at very high precision also contain inner transiting super-Earths interior to the orbits of their \hjs.

\begin{figure}
    \centering
    \includegraphics[width=\columnwidth]{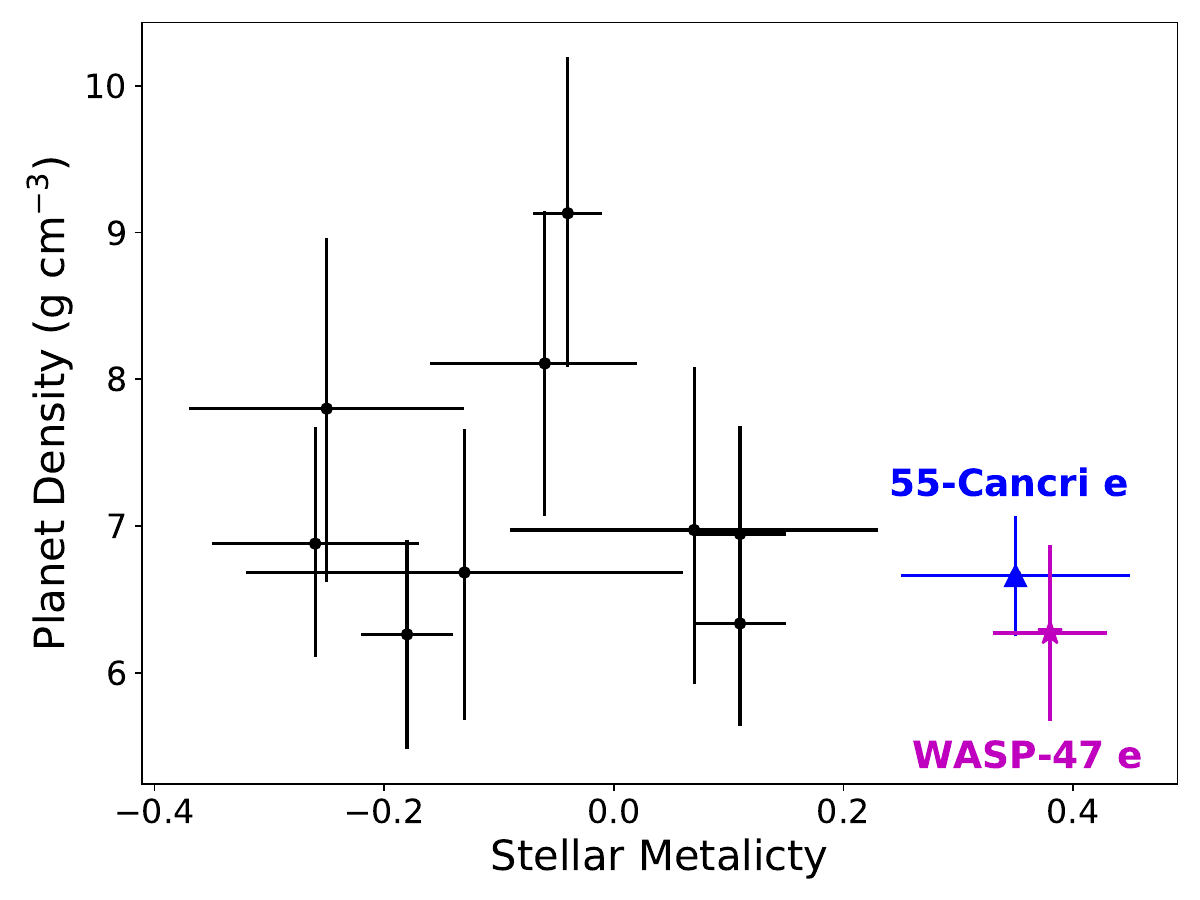}
    \caption{Planet bulk density plotted against stellar metallicity for \Nplanete, \cnce, and the sample of super-Earths from Figure~\ref{fig:MR}.}
    \label{fig:rho_met}
\end{figure}

The two stars \Nstar\ and \cnc\ have metallicities of \feh=\Nmetal\ \citep{vanderburg2017wasp47} and \feh=\Nmetalcnc\ \citep{bourrier201855cnce} respectively. These two stars are significantly more metal rich than the host stars for the other super-Earths in the sample considered in this paper (see Figure~\ref{fig:rho_met}).  While high metallicities are expected for stars which host close-in giant exoplanets \citep{fischervalenti2005pmc}, we find that \Nstar\ and \cnc\ have high metallicities even compared to the metallicity distribution for hot-Jupiter host stars. \citet{osborn2020planetmetalcorr}  investigated this planet-metallicity correlation using a homogeneous sample of 217 \hj\ host stars. We find that \Nstar\ is more metal rich than 96\% of that sample, and \cnc\ is more metal rich than 94\%. It is possible that the high metallicity of the host stars has in some way affected the formation and composition of \Nplanete\ and \cnce. However, we will only be able to confirm this correlation with the discovery of more small planets interior to close-in giant planets.

\section{Conclusions}

We report radial velocity measurements obtained for \Nstar\ using the ESPRESSO spectrograph. We combined our new RV measurements with existing data obtained using the HARPS-N, HIRES and CORALIE spectrographs. Our measurements confirm that \Nplanete\ is a super-Earth with a mass of \Nmasse\,\mearth\ and a bulk density of \Ndensitye\,\gccc. Using these data, we improve the mass measurement precision for the super-Earth \Nplanete\ and the Neptune-sized \Nplanetd\ by 15\% and 13\% respectively compared to the previous best measurements \citep{vanderburg2017wasp47}. Our measured bulk density shows that \Nplanete\ is similar in density, and likely composition, to \cnce.  \Nplanete\ has the second most precisely measured density of any super-Earth, only behind \cnce\ which orbits a much brighter host star.

We find variability in the K2 photometric data which is indicative of a \prot\,=\,32.5$\pm$3.9\,day rotation period for \Nstar. The improved precision of the ESPRESSO data compared to previous data sets allowed us to identify a stellar activity signal in the ESPRESSO radial velocity residuals and activity indicators with a periodicity consistent with the rotation period derived from K2.

With the inclusion of the TESS photometry, we refine the orbital ephemerides of the three inner planets. This refinement is vital for future follow-up observations. As an example, at the start of July 2022 when JWST observations should begin, our updated ephemeris predicts the transit of \Nplanetb\ to occur 20\,minutes later than the older ephemeris of \citet{becker2015wasp47K2}.  Our updated ephemeris has a 1$\sigma$ uncertainty of just 0.57\,minutes, compared to 4.67\,minutes from the older \citet{becker2015wasp47K2} ephemeris. Similarly for \Nplanetd\ our new ephemeris predicts the transit to occur 114\,minutes earlier, with a reduced 1$\sigma$ uncertainty of 35\,minutes compared to 84\,minutes. For \Nplanete\ our new ephemeris predicts the transit to occur just 9.73\,minutes earlier, but with a reduced 1$\sigma$ uncertainty of 25\,minutes compared to 65\,minutes. These refined transit predictions are crucial for properly planning and fully interpreting high value follow-up observations, such as transmission spectroscopy with JWST.

Due to the TESS photometry having a photometric precision of 3040\,ppm-per-minute (c.f. K2 precision of 350\,ppm-per-minute), the TESS data is unable to improve the constraints on the radii of the transiting planets.  For similar reasons the TESS data is not sensitive to the stellar variability seen in the K2 data.

Compared to other well characterized super-Earths, \Nplanete\ and \cnce\ stand out in terms of their low densities given their stellar environments and the presence of a close-orbiting giant planet. It is possible that a different formation mechanism to the majority of \hjs\ was required to form these systems, and this difference in formation could have resulted in \Nplanete\ and \cnce\ having different compositions to other super-Earths.

\begin{table*}  
 \centering  
 \small  
 \caption{Planetary parameters derived in this work in sections \ref{sec:trans_analysis} and \ref{sec:final_analysis}.}
 \label{tab:params}  
 \begin{tabular}{l l l c c }  
 \noalign{\smallskip} \noalign{\smallskip} \hline  \hline \noalign{\smallskip}  
 Parameter  &   Symbol  &  Unit  &  Prior & Value \\   
 \noalign{\smallskip} \hline \noalign{\smallskip} \noalign{\smallskip}  
 \multicolumn{5}{c}{{\bf Planetary parameters}} \\
 \noalign{\smallskip} \noalign{\smallskip} 
 \multicolumn{5}{l}{\it \Nplanete} \\
 \noalign{\smallskip}
 Time of conjunction &  \tc &  BJD (TDB)  &  $\mathcal{N}\left(2457011.34863, 0.00030\right)$ & \Ntce \\
 \noalign{\smallskip} 
 Orbital Period & $P$ & days & $\mathcal{N_U}\left(0.789595, 0.000005, 0., \infty\right)$ & \Nperiode \\
 \noalign{\smallskip}
 RV Semi-Amplitude & $K_e$ & ms$^{-1}$ & $\mathcal{N_U}\left(5., 10., 0., \infty\right)$ & \NKe \\
 \noalign{\smallskip}
 {\bf Radius Ratio} & $R_e / R_*$ & & $\mathcal{U}\left(0., 1.0\right)$ & \Nrratioe \\
 \noalign{\smallskip}
 Planet Mass$^{\dagger}$ & $M_e$ & \mearth & & \Nmasse \\
 \noalign{\smallskip}
 Planet Radius$^{\dagger}$ & $R_e$ & \rearth & & \Nradiuse \\
 \noalign{\smallskip}
 Planet Density$^{\dagger}$ & $\rho_e$ & \gccc & & \Ndensitye \\
 \noalign{\smallskip} \noalign{\smallskip} \noalign{\smallskip}
 \multicolumn{5}{l}{\it \Nplanetb} \\
 \noalign{\smallskip}
 Time of conjunction &  \tc &  BJD (TDB)  &  $\mathcal{N}\left(2457007.932103, 0.000019\right)$ & \Ntcb \\
 \noalign{\smallskip} 
 Orbital Period & $P$ & days & $\mathcal{N_U}\left(4.1591492, 0.0000006, 0., \infty\right)$ & \Nperiodb \\
 \noalign{\smallskip}
 RV Semi-Amplitude & $K_b$ & ms$^{-1}$ & $\mathcal{N_U}\left(150., 50., 0., \infty\right)$ & \NKb \\
 \noalign{\smallskip}
 Radius Ratio & $R_b / R_*$ & & $\mathcal{U}\left(0., 1.0\right)$ & \Nrratiob \\
 \noalign{\smallskip}
 Planet Mass$^{\dagger}$ & $M_b$ & \mearth & & \Nmassb \\
 \noalign{\smallskip}
 Planet Radius$^{\dagger}$ & $R_b$ & \rearth & & \Nradiusb \\
 \noalign{\smallskip}
 Planet Density$^{\dagger}$ & $\rho_b$ & \gccc & & \Ndensityb \\
 \noalign{\smallskip} \noalign{\smallskip} \noalign{\smallskip}
 \multicolumn{5}{l}{\it \Nplanetd} \\
 \noalign{\smallskip}
 Time of conjunction &  \tc &  BJD (TDB)  & $\mathcal{N}\left(2457006.36955, 0.00035\right)$ & \Ntcd \\
 \noalign{\smallskip} 
 Orbital Period & $P$ & days & $\mathcal{N_U}\left(9.03055, 0.0002, 0., \infty\right)$ & \Nperiodd \\
 \noalign{\smallskip}
 RV Semi-Amplitude & $K_d$ & ms$^{-1}$ & $\mathcal{N_U}\left(5., 10., 0., \infty\right)$ & \NKd \\
 \noalign{\smallskip}
 Radius Ratio & $R_d / R_*$ & & $\mathcal{U}\left(0., 1.0\right)$ & \Nrratiod \\
 \noalign{\smallskip}
 Planet Mass$^{\dagger}$ & $M_d$ & \mearth & & \Nmassd \\
 \noalign{\smallskip}
 Planet Radius$^{\dagger}$ & $R_d$ & \rearth & & \Nradiusd \\
 \noalign{\smallskip}
 Planet Density$^{\dagger}$ & $\rho_d$ & \gccc & & \Ndensityd \\
 \noalign{\smallskip}
 Orbital Eccentricity & e$_d$ & & $\mathcal{N_U}\left(0., 0.014, 0., 1.\right)$ & \Neccd \\
 \noalign{\smallskip}
 Argument of Periastron & $\omega_d$ & deg & $\mathcal{U}\left(-180., 180.\right)$ & \Nomegad \\
 \noalign{\smallskip} \noalign{\smallskip} \noalign{\smallskip}
 \multicolumn{5}{l}{\it \Nplanetc} \\
 \noalign{\smallskip}
 Time of conjunction &  \tc &  BJD (TDB)  &  $\mathcal{N}\left(2457763.4, 20.\right)$ & \Ntcc \\
 \noalign{\smallskip} 
 Orbital Period & $P$ & days & $\mathcal{N_U}\left(590, 15., 0., \infty\right)$ & \Nperiodc \\
 \noalign{\smallskip}
 RV Semi-Amplitude & $K_c$ & ms$^{-1}$ & $\mathcal{N_U}\left(30., 15., 0., \infty\right)$ & \NKc \\
 \noalign{\smallskip}
 Planet Minimum Mass$^{\dagger}$ & $M_c\,\sin\,i$ & \mearth & & \Nmassc \\
 \noalign{\smallskip}
 Orbital Eccentricity & e$_c$ & & $\mathcal{U}\left(0., 1.\right)$ & \Neccc \\
 \noalign{\smallskip}
 Argument of Periastron & $\omega_c$ & deg & $\mathcal{U}\left(-180., 180.\right)$ & \Nomegac \\
 \noalign{\smallskip} \noalign{\smallskip} \noalign{\smallskip}
 \hline \hline
\end{tabular}  
 \begin{list}{}{}
 \item[$\dagger$ - Parameter derived from fitted parameters]
 \end{list} 
 \end{table*}
 
\begin{table*}
 \centering
 \small  
 \caption{Additional parameters from the sampling in Section~\ref{sec:final_analysis}.}
 \label{tab:params2}  
 \begin{tabular}{l l c c }  
 \noalign{\smallskip} \noalign{\smallskip} \hline  \hline \noalign{\smallskip}  
 Parameter  &  Unit  &  Prior & Value \\   
 \noalign{\smallskip} \hline \noalign{\smallskip} \noalign{\smallskip}
 \multicolumn{4}{c}{{\bf Instrumental Parameters}} \\
 \noalign{\smallskip} \noalign{\smallskip}
 $\gamma_{\rm ESP}$ & ms$^{-1}$ & $\mathcal{U}\left(-27350., -26850.\right)$ & \Ngamesp \\
 $\gamma_{\rm HARPS}$ & ms$^{-1}$ & $\mathcal{U}\left(-27350., -26850.\right)$ & \Ngamharps \\
 $\gamma_{\rm HIRES}$ & ms$^{-1}$ & $\mathcal{U}\left(-250., 250.\right)$ & \Ngamhires \\
 $\gamma_{\rm COR; A}$\,$^{\dagger}$ & ms$^{-1}$ & $\mathcal{U}\left(-27500., -26500.\right)$ & \NgamcorA \\
 $\gamma_{\rm COR; B}$\,$^{\dagger}$ & ms$^{-1}$ & $\mathcal{U}\left(-27500., -26500.\right)$ & \NgamcorB \\
 $\gamma_{\rm ESP FWHM}$ & ms$^{-1}$ & $\mathcal{U}\left(7600., 7700.\right)$ & \Ngamfwhm \\
 $\log_{10}\,\sigma_{\rm ESP}$ & $\log\,$ms$^{-1}$ & $\mathcal{U}\left(-5., 1.\right)$ & \Nlogjitesp \\
 $\log_{10}\,\sigma_{\rm HARPS}$ & $\log\,$ms$^{-1}$ & $\mathcal{U}\left(-5., 1.\right)$ & \Nlogjitharps \\
 $\log_{10}\,\sigma_{\rm HIRES}$ & $\log\,$ms$^{-1}$ & $\mathcal{U}\left(-5., 1.\right)$ & \Nlogjithires \\
 $\log_{10}\,\sigma_{\rm COR; A}$\,$^{\dagger}$ & $\log\,$ms$^{-1}$ & $\mathcal{U}\left(-7., 2.\right)$ & \NlogjitcorA \\
 $\log_{10}\,\sigma_{\rm COR; B}$\,$^{\dagger}$ & $\log\,$ms$^{-1}$ & $\mathcal{U}\left(-7., 2.\right)$ & \NlogjitcorB \\
 $\log_{10}\,\sigma_{\rm ESP FWHM}$ & $\log\,$ms$^{-1}$ & $\mathcal{U}\left(-7., 1.\right)$ & \Nlogjitfwhm \\
 \multicolumn{4}{c}{{\bf GP Hyperparameters}} \\
 \noalign{\smallskip} \noalign{\smallskip} 
 \prot & days & $\mathcal{N_U}\left(35, 10, 28, \infty\right)$ & \Nprot \\
 \noalign{\smallskip}
 $\ln\,Q$ & & $\mathcal{U}\left(1.15, 15\right)$ & \Nlogq \\
 \noalign{\smallskip}
 $\ln\,\Delta\,Q$ & & $\mathcal{U}\left(-7, 10\right)$ & \Nlogdq \\
 \noalign{\smallskip}
 $f$ & & $\mathcal{U}\left(0, 1\right)$ & \Nf \\
 \noalign{\smallskip}
 $\sigma_{\rm GP, ESP}$ & & $\mathcal{U}\left(0, 20\right)$ & \Nsiggpesp \\
 \noalign{\smallskip}
 $\sigma_{\rm GP, ESP FWHM}$ & & $\mathcal{U}\left(0, 100\right)$ & \Nsiggpfwhm \\
 \noalign{\smallskip} \hline \hline \noalign{\smallskip} \noalign{\smallskip} 
\end{tabular}
\begin{list}{}{}
 \item[$\dagger$ - COR;A and COR;B refer to the sets of CORALIE data taken before and after the spectrograph was updated in November 2014]
 \end{list}
 \end{table*}

\section*{Acknowledgements}

This research made use of \textsc{exoplanet} \citep{exoplanet:exoplanet} and its
dependencies: \textsc{celerite2} \citep{celerite2:foremanmackey17, celerite2:foremanmackey18}; \textsc{astropy} \citep{exoplanet:astropy13, exoplanet:astropy18}, \textsc{PyMC3} \citep{exoplanet:pymc3}, and \textsc{theano} \citep{exoplanet:theano}. This research has made use of the NASA Exoplanet Archive, which is operated by the California Institute of Technology, under contract with the National Aeronautics and Space Administration under the Exoplanet Exploration Program. Based on observations made with ESPRESSO on the Very Large Telescope under ESO observing program 0103.C-0422. This paper includes data collected by the Kepler mission and obtained from the MAST data archive at the Space Telescope Science Institute (STScI). Funding for the Kepler mission is provided by the NASA Science Mission Directorate. STScI is operated by the Association of Universities for Research in Astronomy, Inc., under NASA contract NAS 5–26555. 

\section*{Data Availability}

The observations made with ESPRESSO (program 0103.C-0422) are publicly available through the ESO archive (\href{http://archive.eso.org/}{http://archive.eso.org/}) and the reduced radial-velocities are available in Table~\ref{tab:espresso_rvs}. The K2-SFF long cadence photometry is available as a High Level Science Product through the MAST database (\href{https://archive.stsci.edu/hlsp/k2sff}{https://archive.stsci.edu/hlsp/k2sff}), and the short candence light curve is available for download from \href{http://www.cfa.harvard.edu/~avanderb/wasp47sc.csv}{http://www.cfa.harvard.edu/~avanderb/wasp47sc.csv}. The archival radial velocity measurements used were accessed and are available through the online versions of the corresponding publications: HARPS-N \citep{vanderburg2017wasp47}; HIRES \citep{sinukoff2017wasp47hires}; CORALIE \citep{neveuVanMalle2016wasp47Coralie}; PFS \citep{dai2015wasp47PFS}.


\bibliography{wasp47}{}

\begin{thebibliography}{}
\expandafter\ifx\csname natexlab\endcsname\relax\def\natexlab#1{#1}\fi
\providecommand{\url}[1]{\href{#1}{#1}}
\providecommand{\dodoi}[1]{doi:~\href{http://doi.org/#1}{\nolinkurl{#1}}}
\providecommand{\doeprint}[1]{\href{http://ascl.net/#1}{\nolinkurl{http://ascl.net/#1}}}
\providecommand{\doarXiv}[1]{\href{https://arxiv.org/abs/#1}{\nolinkurl{https://arxiv.org/abs/#1}}}

\bibitem[{{Almenara} {et~al.}(2016){Almenara}, {D{\'\i}az}, {Bonfils}, \&
  {Udry}}]{almenara2016wasp47_dynmass}
{Almenara}, J.~M., {D{\'\i}az}, R.~F., {Bonfils}, X., \& {Udry}, S. 2016, \aap,
  595, L5, \dodoi{10.1051/0004-6361/201629770}

\bibitem[{{Angelo} \& {Hu}(2017)}]{angelo201755cancrispitzeratmos}
{Angelo}, I., \& {Hu}, R. 2017, \aj, 154, 232, \dodoi{10.3847/1538-3881/aa9278}

\bibitem[{{Astropy Collaboration} {et~al.}(2013){Astropy Collaboration},
  {Robitaille}, {Tollerud}, {Greenfield}, {Droettboom}, {Bray}, {Aldcroft},
  {Davis}, {Ginsburg}, {Price-Whelan}, {Kerzendorf}, {Conley}, {Crighton},
  {Barbary}, {Muna}, {Ferguson}, {Grollier}, {Parikh}, {Nair}, {Unther},
  {Deil}, {Woillez}, {Conseil}, {Kramer}, {Turner}, {Singer}, {Fox}, {Weaver},
  {Zabalza}, {Edwards}, {Azalee Bostroem}, {Burke}, {Casey}, {Crawford},
  {Dencheva}, {Ely}, {Jenness}, {Labrie}, {Lim}, {Pierfederici}, {Pontzen},
  {Ptak}, {Refsdal}, {Servillat}, \& {Streicher}}]{exoplanet:astropy13}
{Astropy Collaboration}, {Robitaille}, T.~P., {Tollerud}, E.~J., {et~al.} 2013,
  \aap, 558, A33, \dodoi{10.1051/0004-6361/201322068}

\bibitem[{{Astropy Collaboration} {et~al.}(2018){Astropy Collaboration},
  {Price-Whelan}, {Sip{\H o}cz}, {G{\"u}nther}, {Lim}, {Crawford}, {Conseil},
  {Shupe}, {Craig}, {Dencheva}, {Ginsburg}, {VanderPlas}, {Bradley},
  {P{\'e}rez-Su{\'a}rez}, {de Val-Borro}, {Aldcroft}, {Cruz}, {Robitaille},
  {Tollerud}, {Ardelean}, {Babej}, {Bach}, {Bachetti}, {Bakanov}, {Bamford},
  {Barentsen}, {Barmby}, {Baumbach}, {Berry}, {Biscani}, {Boquien}, {Bostroem},
  {Bouma}, {Brammer}, {Bray}, {Breytenbach}, {Buddelmeijer}, {Burke},
  {Calderone}, {Cano Rodr{\'{\i}}guez}, {Cara}, {Cardoso}, {Cheedella},
  {Copin}, {Corrales}, {Crichton}, {D'Avella}, {Deil}, {Depagne}, {Dietrich},
  {Donath}, {Droettboom}, {Earl}, {Erben}, {Fabbro}, {Ferreira}, {Finethy},
  {Fox}, {Garrison}, {Gibbons}, {Goldstein}, {Gommers}, {Greco}, {Greenfield},
  {Groener}, {Grollier}, {Hagen}, {Hirst}, {Homeier}, {Horton}, {Hosseinzadeh},
  {Hu}, {Hunkeler}, {Ivezi{\'c}}, {Jain}, {Jenness}, {Kanarek}, {Kendrew},
  {Kern}, {Kerzendorf}, {Khvalko}, {King}, {Kirkby}, {Kulkarni}, {Kumar},
  {Lee}, {Lenz}, {Littlefair}, {Ma}, {Macleod}, {Mastropietro}, {McCully},
  {Montagnac}, {Morris}, {Mueller}, {Mumford}, {Muna}, {Murphy}, {Nelson},
  {Nguyen}, {Ninan}, {N{\"o}the}, {Ogaz}, {Oh}, {Parejko}, {Parley}, {Pascual},
  {Patil}, {Patil}, {Plunkett}, {Prochaska}, {Rastogi}, {Reddy Janga},
  {Sabater}, {Sakurikar}, {Seifert}, {Sherbert}, {Sherwood-Taylor}, {Shih},
  {Sick}, {Silbiger}, {Singanamalla}, {Singer}, {Sladen}, {Sooley},
  {Sornarajah}, {Streicher}, {Teuben}, {Thomas}, {Tremblay}, {Turner},
  {Terr{\'o}n}, {van Kerkwijk}, {de la Vega}, {Watkins}, {Weaver}, {Whitmore},
  {Woillez}, {Zabalza}, \& {Astropy Contributors}}]{exoplanet:astropy18}
{Astropy Collaboration}, {Price-Whelan}, A.~M., {Sip{\H o}cz}, B.~M., {et~al.}
  2018, \aj, 156, 123, \dodoi{10.3847/1538-3881/aabc4f}

\bibitem[{{Barnes}(2007)}]{barnes2007fieldstargyrochron}
{Barnes}, S.~A. 2007, \apj, 669, 1167, \dodoi{10.1086/519295}

\bibitem[{{Becker} {et~al.}(2015){Becker}, {Vanderburg}, {Adams}, {Rappaport},
  \& {Schwengeler}}]{becker2015wasp47K2}
{Becker}, J.~C., {Vanderburg}, A., {Adams}, F.~C., {Rappaport}, S.~A., \&
  {Schwengeler}, H.~M. 2015, \apjl, 812, L18,
  \dodoi{10.1088/2041-8205/812/2/L18}

\bibitem[{{Boisse} {et~al.}(2011){Boisse}, {Bouchy}, {H{\'e}brard}, {Bonfils},
  {Santos}, \& {Vauclair}}]{boisse2011stellaractivity}
{Boisse}, I., {Bouchy}, F., {H{\'e}brard}, G., {et~al.} 2011, \aap, 528, A4,
  \dodoi{10.1051/0004-6361/201014354}

\bibitem[{{Borucki} {et~al.}(2010){Borucki}, {Koch}, {Basri}, {Batalha},
  {Brown}, {Caldwell}, {Caldwell}, {Christensen-Dalsgaard}, {Cochran},
  {DeVore}, {Dunham}, {Dupree}, {Gautier}, {Geary}, {Gilliland}, {Gould},
  {Howell}, {Jenkins}, {Kondo}, {Latham}, {Marcy}, {Meibom}, {Kjeldsen},
  {Lissauer}, {Monet}, {Morrison}, {Sasselov}, {Tarter}, {Boss}, {Brownlee},
  {Owen}, {Buzasi}, {Charbonneau}, {Doyle}, {Fortney}, {Ford}, {Holman},
  {Seager}, {Steffen}, {Welsh}, {Rowe}, {Anderson}, {Buchhave}, {Ciardi},
  {Walkowicz}, {Sherry}, {Horch}, {Isaacson}, {Everett}, {Fischer}, {Torres},
  {Johnson}, {Endl}, {MacQueen}, {Bryson}, {Dotson}, {Haas}, {Kolodziejczak},
  {Van Cleve}, {Chandrasekaran}, {Twicken}, {Quintana}, {Clarke}, {Allen},
  {Li}, {Wu}, {Tenenbaum}, {Verner}, {Bruhweiler}, {Barnes}, \&
  {Prsa}}]{borucki2010kepler}
{Borucki}, W.~J., {Koch}, D., {Basri}, G., {et~al.} 2010, Science, 327, 977,
  \dodoi{10.1126/science.1185402}

\bibitem[{{Bourrier} {et~al.}(2018){Bourrier}, {Dumusque}, {Dorn}, {Henry},
  {Astudillo-Defru}, {Rey}, {Benneke}, {H{\'e}brard}, {Lovis}, {Demory},
  {Moutou}, \& {Ehrenreich}}]{bourrier201855cnce}
{Bourrier}, V., {Dumusque}, X., {Dorn}, C., {et~al.} 2018, \aap, 619, A1,
  \dodoi{10.1051/0004-6361/201833154}

\bibitem[{{Ca{\~n}as} {et~al.}(2019){Ca{\~n}as}, {Wang}, {Mahadevan}, {Bender},
  {De Lee}, {Fleming}, {Garc{\'\i}a-Hern{\'a}ndez}, {Hearty}, {Majewski},
  {Roman-Lopes}, {Schneider}, \& {Stassun}}]{canas2019kep730}
{Ca{\~n}as}, C.~I., {Wang}, S., {Mahadevan}, S., {et~al.} 2019, \apjl, 870,
  L17, \dodoi{10.3847/2041-8213/aafa1e}

\bibitem[{{Chatterjee} {et~al.}(2008){Chatterjee}, {Ford}, {Matsumura}, \&
  {Rasio}}]{chatterjee2008planetscattering}
{Chatterjee}, S., {Ford}, E.~B., {Matsumura}, S., \& {Rasio}, F.~A. 2008, \apj,
  686, 580, \dodoi{10.1086/590227}

\bibitem[{{Clarke}(2003)}]{clarke2003photvariability}
{Clarke}, D. 2003, \aap, 407, 1029, \dodoi{10.1051/0004-6361:20030901}

\bibitem[{{Dai} {et~al.}(2015){Dai}, {Winn}, {Arriagada}, {Butler}, {Crane},
  {Johnson}, {Shectman}, {Teske}, {Thompson}, {Vanderburg}, \&
  {Wittenmyer}}]{dai2015wasp47PFS}
{Dai}, F., {Winn}, J.~N., {Arriagada}, P., {et~al.} 2015, \apjl, 813, L9,
  \dodoi{10.1088/2041-8205/813/1/L9}

\bibitem[{{Dawson} \& {Johnson}(2018)}]{dawsonjohnson2018hotjupiters}
{Dawson}, R.~I., \& {Johnson}, J.~A. 2018, \araa, 56, 175,
  \dodoi{10.1146/annurev-astro-081817-051853}

\bibitem[{{Demory} {et~al.}(2012){Demory}, {Gillon}, {Seager}, {Benneke},
  {Deming}, \& {Jackson}}]{demory201255cncemission}
{Demory}, B.-O., {Gillon}, M., {Seager}, S., {et~al.} 2012, \apjl, 751, L28,
  \dodoi{10.1088/2041-8205/751/2/L28}

\bibitem[{{Donahue} {et~al.}(1997){Donahue}, {Dobson}, \&
  {Baliunas}}]{donahue1997}
{Donahue}, R.~A., {Dobson}, A.~K., \& {Baliunas}, S.~L. 1997, \solphys, 171,
  191, \dodoi{10.1023/A:1004902307998}

\bibitem[{{Dorn} {et~al.}(2019){Dorn}, {Harrison}, {Bonsor}, \&
  {Hands}}]{dorn2019superearthformation}
{Dorn}, C., {Harrison}, J.~H.~D., {Bonsor}, A., \& {Hands}, T.~O. 2019, \mnras,
  484, 712, \dodoi{10.1093/mnras/sty3435}

\bibitem[{{Dreizler} {et~al.}(2020){Dreizler}, {Crossfield}, {Kossakowski},
  {Plavchan}, {Jeffers}, {Kemmer}, {Luque}, {Espinoza}, {Pall{\'e}}, {Stassun},
  {Matthews}, {Cale}, {Caballero}, {Schlecker}, {Lillo-Box}, {Zechmeister},
  {Lalitha}, {Reiners}, {Soubkiou}, {Bitsch}, {Zapatero Osorio}, {Chaturvedi},
  {Hatzes}, {Ricker}, {Vanderspek}, {Latham}, {Seager}, {Winn}, {Jenkins},
  {Aceituno}, {Amado}, {Barkaoui}, {Barbieri}, {Batalha}, {Bauer}, {Benneke},
  {Benkhaldoun}, {Beichman}, {Berberian}, {Burt}, {Butler}, {Caldwell},
  {Chintada}, {Chontos}, {Christiansen}, {Ciardi}, {Cifuentes}, {Collins},
  {Collins}, {Combs}, {Cort{\'e}s-Contreras}, {Crane}, {Daylan}, {Dragomir},
  {Esparza-Borges}, {Evans}, {Feng}, {Flowers}, {Fukui}, {Fulton}, {Furlan},
  {Gaidos}, {Geneser}, {Giacalone}, {Gillon}, {Gonzales}, {Gorjian}, {Hellier},
  {Hidalgo}, {Howard}, {Howell}, {Huber}, {Isaacson}, {Jehin}, {Jensen},
  {Kaminski}, {Kane}, {Kawauchi}, {Kielkopf}, {Klahr}, {Kosiarek}, {Kreidberg},
  {K{\"u}rster}, {Lafarga}, {Livingston}, {Louie}, {Mann}, {Madrigal-Aguado},
  {Matson}, {Mocnik}, {Morales}, {Muirhead}, {Murgas}, {Nandakumar}, {Narita},
  {Nowak}, {Oshagh}, {Parviainen}, {Passegger}, {Pollacco}, {Pozuelos},
  {Quirrenbach}, {Reefe}, {Ribas}, {Robertson}, {Rodr{\'\i}guez-L{\'o}pez},
  {Rose}, {Roy}, {Schweitzer}, {Schlieder}, {Shectman}, {Tanner},
  {{\c{S}}enavc{\i}}, {Teske}, {Twicken}, {Villasenor}, {Wang}, {Weiss},
  {Wittrock}, {Y{\i}lmaz}, \& {Zohrabi}}]{dreizler2020lp71447}
{Dreizler}, S., {Crossfield}, I.~J.~M., {Kossakowski}, D., {et~al.} 2020, \aap,
  644, A127, \dodoi{10.1051/0004-6361/202038016}

\bibitem[{{Fischer} \& {Valenti}(2005)}]{fischervalenti2005pmc}
{Fischer}, D.~A., \& {Valenti}, J. 2005, \apj, 622, 1102,
  \dodoi{10.1086/428383}

\bibitem[{{Fogg} \& {Nelson}(2005)}]{foggnelson2005}
{Fogg}, M.~J., \& {Nelson}, R.~P. 2005, \aap, 441, 791,
  \dodoi{10.1051/0004-6361:20053453}

\bibitem[{{Fogg} \& {Nelson}(2007)}]{foggnelson2007typeImigration}
---. 2007, \aap, 472, 1003, \dodoi{10.1051/0004-6361:20077950}

\bibitem[{{Foreman-Mackey}(2018{\natexlab{a}})}]{celerite2}
{Foreman-Mackey}, D. 2018{\natexlab{a}}, Research Notes of the American
  Astronomical Society, 2, 31, \dodoi{10.3847/2515-5172/aaaf6c}

\bibitem[{{Foreman-Mackey}(2018{\natexlab{b}})}]{celerite2:foremanmackey18}
---. 2018{\natexlab{b}}, Research Notes of the American Astronomical Society,
  2, 31, \dodoi{10.3847/2515-5172/aaaf6c}

\bibitem[{{Foreman-Mackey} {et~al.}(2017{\natexlab{a}}){Foreman-Mackey},
  {Agol}, {Ambikasaran}, \& {Angus}}]{celerite1}
{Foreman-Mackey}, D., {Agol}, E., {Ambikasaran}, S., \& {Angus}, R.
  2017{\natexlab{a}}, \aj, 154, 220, \dodoi{10.3847/1538-3881/aa9332}

\bibitem[{{Foreman-Mackey} {et~al.}(2017{\natexlab{b}}){Foreman-Mackey},
  {Agol}, {Ambikasaran}, \& {Angus}}]{celerite2:foremanmackey17}
---. 2017{\natexlab{b}}, \aj, 154, 220, \dodoi{10.3847/1538-3881/aa9332}

\bibitem[{{Foreman-Mackey} {et~al.}(2013){Foreman-Mackey}, {Hogg}, {Lang}, \&
  {Goodman}}]{foremanmackey2013emcee}
{Foreman-Mackey}, D., {Hogg}, D.~W., {Lang}, D., \& {Goodman}, J. 2013, \pasp,
  125, 306, \dodoi{10.1086/670067}

\bibitem[{Foreman-Mackey {et~al.}(2020)Foreman-Mackey, Luger, Czekala, Agol,
  Price-Whelan, \& Barclay}]{exoplanet:exoplanet}
Foreman-Mackey, D., Luger, R., Czekala, I., {et~al.} 2020,
  exoplanet-dev/exoplanet v0.3.2, \dodoi{10.5281/zenodo.1998447}

\bibitem[{{Freudling} {et~al.}(2013){Freudling}, {Romaniello}, {Bramich},
  {Ballester}, {Forchi}, {Garc{\'{\i}}a-Dabl{\'o}}, {Moehler}, \&
  {Neeser}}]{esoreflex2013}
{Freudling}, W., {Romaniello}, M., {Bramich}, D.~M., {et~al.} 2013, \aap, 559,
  A96, \dodoi{10.1051/0004-6361/201322494}

\bibitem[{{Gaia Collaboration} {et~al.}(2021){Gaia Collaboration}, {Brown},
  {Vallenari}, {Prusti}, {de Bruijne}, {Babusiaux}, {Biermann}, {Creevey},
  {Evans}, {Eyer}, {Hutton}, {Jansen}, {Jordi}, {Klioner}, {Lammers},
  {Lindegren}, {Luri}, {Mignard}, {Panem}, {Pourbaix}, {Randich}, {Sartoretti},
  {Soubiran}, {Walton}, {Arenou}, {Bailer-Jones}, {Bastian}, {Cropper},
  {Drimmel}, {Katz}, {Lattanzi}, {van Leeuwen}, {Bakker}, {Cacciari},
  {Casta{\~n}eda}, {De Angeli}, {Ducourant}, {Fabricius}, {Fouesneau},
  {Fr{\'e}mat}, {Guerra}, {Guerrier}, {Guiraud}, {Jean-Antoine Piccolo},
  {Masana}, {Messineo}, {Mowlavi}, {Nicolas}, {Nienartowicz}, {Pailler},
  {Panuzzo}, {Riclet}, {Roux}, {Seabroke}, {Sordo}, {Tanga}, {Th{\'e}venin},
  {Gracia-Abril}, {Portell}, {Teyssier}, {Altmann}, {Andrae}, {Bellas-Velidis},
  {Benson}, {Berthier}, {Blomme}, {Brugaletta}, {Burgess}, {Busso}, {Carry},
  {Cellino}, {Cheek}, {Clementini}, {Damerdji}, {Davidson}, {Delchambre},
  {Dell'Oro}, {Fern{\'a}ndez-Hern{\'a}ndez}, {Galluccio}, {Garc{\'\i}a-Lario},
  {Garcia-Reinaldos}, {Gonz{\'a}lez-N{\'u}{\~n}ez}, {Gosset}, {Haigron},
  {Halbwachs}, {Hambly}, {Harrison}, {Hatzidimitriou}, {Heiter},
  {Hern{\'a}ndez}, {Hestroffer}, {Hodgkin}, {Holl}, {Jan{\ss}en}, {Jevardat de
  Fombelle}, {Jordan}, {Krone-Martins}, {Lanzafame}, {L{\"o}ffler}, {Lorca},
  {Manteiga}, {Marchal}, {Marrese}, {Moitinho}, {Mora}, {Muinonen}, {Osborne},
  {Pancino}, {Pauwels}, {Petit}, {Recio-Blanco}, {Richards}, {Riello},
  {Rimoldini}, {Robin}, {Roegiers}, {Rybizki}, {Sarro}, {Siopis}, {Smith},
  {Sozzetti}, {Ulla}, {Utrilla}, {van Leeuwen}, {van Reeven}, {Abbas}, {Abreu
  Aramburu}, {Accart}, {Aerts}, {Aguado}, {Ajaj}, {Altavilla}, {{\'A}lvarez},
  {{\'A}lvarez Cid-Fuentes}, {Alves}, {Anderson}, {Anglada Varela}, {Antoja},
  {Audard}, {Baines}, {Baker}, {Balaguer-N{\'u}{\~n}ez}, {Balbinot}, {Balog},
  {Barache}, {Barbato}, {Barros}, {Barstow}, {Bartolom{\'e}}, {Bassilana},
  {Bauchet}, {Baudesson-Stella}, {Becciani}, {Bellazzini}, {Bernet}, {Bertone},
  {Bianchi}, {Blanco-Cuaresma}, {Boch}, {Bombrun}, {Bossini}, {Bouquillon},
  {Bragaglia}, {Bramante}, {Breedt}, {Bressan}, {Brouillet}, {Bucciarelli},
  {Burlacu}, {Busonero}, {Butkevich}, {Buzzi}, {Caffau}, {Cancelliere},
  {C{\'a}novas}, {Cantat-Gaudin}, {Carballo}, {Carlucci}, {Carnerero},
  {Carrasco}, {Casamiquela}, {Castellani}, {Castro-Ginard}, {Castro Sampol},
  {Chaoul}, {Charlot}, {Chemin}, {Chiavassa}, {Cioni}, {Comoretto}, {Cooper},
  {Cornez}, {Cowell}, {Crifo}, {Crosta}, {Crowley}, {Dafonte}, {Dapergolas},
  {David}, {David}, {de Laverny}, {De Luise}, {De March}, {De Ridder}, {de
  Souza}, {de Teodoro}, {de Torres}, {del Peloso}, {del Pozo}, {Delbo},
  {Delgado}, {Delgado}, {Delisle}, {Di Matteo}, {Diakite}, {Diener},
  {Distefano}, {Dolding}, {Eappachen}, {Edvardsson}, {Enke}, {Esquej}, {Fabre},
  {Fabrizio}, {Faigler}, {Fedorets}, {Fernique}, {Fienga}, {Figueras},
  {Fouron}, {Fragkoudi}, {Fraile}, {Franke}, {Gai}, {Garabato},
  {Garcia-Gutierrez}, {Garc{\'\i}a-Torres}, {Garofalo}, {Gavras}, {Gerlach},
  {Geyer}, {Giacobbe}, {Gilmore}, {Girona}, {Giuffrida}, {Gomel}, {Gomez},
  {Gonzalez-Santamaria}, {Gonz{\'a}lez-Vidal}, {Granvik},
  {Guti{\'e}rrez-S{\'a}nchez}, {Guy}, {Hauser}, {Haywood}, {Helmi}, {Hidalgo},
  {Hilger}, {H{\l}adczuk}, {Hobbs}, {Holland}, {Huckle}, {Jasniewicz},
  {Jonker}, {Juaristi Campillo}, {Julbe}, {Karbevska}, {Kervella}, {Khanna},
  {Kochoska}, {Kontizas}, {Kordopatis}, {Korn}, {Kostrzewa-Rutkowska},
  {Kruszy{\'n}ska}, {Lambert}, {Lanza}, {Lasne}, {Le Campion}, {Le Fustec},
  {Lebreton}, {Lebzelter}, {Leccia}, {Leclerc}, {Lecoeur-Taibi}, {Liao},
  {Licata}, {Lindstr{\o}m}, {Lister}, {Livanou}, {Lobel}, {Madrero Pardo},
  {Managau}, {Mann}, {Marchant}, {Marconi}, {Marcos Santos}, {Marinoni},
  {Marocco}, {Marshall}, {Martin Polo}, {Mart{\'\i}n-Fleitas}, {Masip},
  {Massari}, {Mastrobuono-Battisti}, {Mazeh}, {McMillan}, {Messina},
  {Michalik}, {Millar}, {Mints}, {Molina}, {Molinaro}, {Moln{\'a}r},
  {Montegriffo}, {Mor}, {Morbidelli}, {Morel}, {Morris}, {Mulone}, {Munoz},
  {Muraveva}, {Murphy}, {Musella}, {Noval}, {Ord{\'e}novic}, {Orr{\`u}},
  {Osinde}, {Pagani}, {Pagano}, {Palaversa}, {Palicio}, {Panahi}, {Pawlak},
  {Pe{\~n}alosa Esteller}, {Penttil{\"a}}, {Piersimoni}, {Pineau}, {Plachy},
  {Plum}, {Poggio}, {Poretti}, {Poujoulet}, {Pr{\v{s}}a}, {Pulone}, {Racero},
  {Ragaini}, {Rainer}, {Raiteri}, {Rambaux}, {Ramos}, {Ramos-Lerate}, {Re
  Fiorentin}, {Regibo}, {Reyl{\'e}}, {Ripepi}, {Riva}, {Rixon}, {Robichon},
  {Robin}, {Roelens}, {Rohrbasser}, {Romero-G{\'o}mez}, {Rowell}, {Royer},
  {Rybicki}, {Sadowski}, {Sagrist{\`a} Sell{\'e}s}, {Sahlmann}, {Salgado},
  {Salguero}, {Samaras}, {Sanchez Gimenez}, {Sanna}, {Santove{\~n}a},
  {Sarasso}, {Schultheis}, {Sciacca}, {Segol}, {Segovia}, {S{\'e}gransan},
  {Semeux}, {Shahaf}, {Siddiqui}, {Siebert}, {Siltala}, {Slezak}, {Smart},
  {Solano}, {Solitro}, {Souami}, {Souchay}, {Spagna}, {Spoto}, {Steele},
  {Steidelm{\"u}ller}, {Stephenson}, {S{\"u}veges}, {Szabados}, {Szegedi-Elek},
  {Taris}, {Tauran}, {Taylor}, {Teixeira}, {Thuillot}, {Tonello}, {Torra},
  {Torra}, {Turon}, {Unger}, {Vaillant}, {van Dillen}, {Vanel}, {Vecchiato},
  {Viala}, {Vicente}, {Voutsinas}, {Weiler}, {Wevers}, {Wyrzykowski}, {Yoldas},
  {Yvard}, {Zhao}, {Zorec}, {Zucker}, {Zurbach}, \& {Zwitter}}]{gaia2021edr3}
{Gaia Collaboration}, {Brown}, A.~G.~A., {Vallenari}, A., {et~al.} 2021, \aap,
  649, A1, \dodoi{10.1051/0004-6361/202039657}

\bibitem[{{Gelman} \& {Rubin}(1992)}]{gelmanrubin92}
{Gelman}, A., \& {Rubin}, D.~B. 1992, Statistical Science, 7, 457,
  \dodoi{10.1214/ss/1177011136}

\bibitem[{{Hellier} {et~al.}(2012){Hellier}, {Anderson}, {Collier Cameron},
  {Doyle}, {Fumel}, {Gillon}, {Jehin}, {Lendl}, {Maxted}, {Pepe}, {Pollacco},
  {Queloz}, {S{\'e}gransan}, {Smalley}, {Smith}, {Southworth}, {Triaud},
  {Udry}, \& {West}}]{hellier2012wasp47}
{Hellier}, C., {Anderson}, D.~R., {Collier Cameron}, A., {et~al.} 2012, \mnras,
  426, 739, \dodoi{10.1111/j.1365-2966.2012.21780.x}

\bibitem[{{Howell} {et~al.}(2014){Howell}, {Sobeck}, {Haas}, {Still},
  {Barclay}, {Mullally}, {Troeltzsch}, {Aigrain}, {Bryson}, {Caldwell},
  {Chaplin}, {Cochran}, {Huber}, {Marcy}, {Miglio}, {Najita}, {Smith},
  {Twicken}, \& {Fortney}}]{howell2014K2}
{Howell}, S.~B., {Sobeck}, C., {Haas}, M., {et~al.} 2014, \pasp, 126, 398,
  \dodoi{10.1086/676406}

\bibitem[{{Huang} {et~al.}(2016){Huang}, {Wu}, \&
  {Triaud}}]{huang2016lonelyHJs}
{Huang}, C., {Wu}, Y., \& {Triaud}, A. H.~M.~J. 2016, \apj, 825, 98,
  \dodoi{10.3847/0004-637X/825/2/98}

\bibitem[{{Huang} {et~al.}(2020){Huang}, {Quinn}, {Vanderburg}, {Becker},
  {Rodriguez}, {Pozuelos}, {Gandolfi}, {Zhou}, {Mann}, {Collins}, {Crossfield},
  {Barkaoui}, {Collins}, {Fridlund}, {Gillon}, {Gonzales}, {G{\"u}nther},
  {Henry}, {Howell}, {James}, {Jao}, {Jehin}, {Jensen}, {Kane}, {Lissauer},
  {Matthews}, {Matson}, {Paredes}, {Schlieder}, {Stassun}, {Shporer}, {Sha},
  {Tan}, {Georgieva}, {Mathur}, {Palle}, {Persson}, {Van Eylen}, {Ricker},
  {Vanderspek}, {Latham}, {Winn}, {Seager}, {Jenkins}, {Burke}, {Goeke},
  {Rinehart}, {Rose}, {Ting}, {Torres}, \& {Wong}}]{huang2020toi1130}
{Huang}, C.~X., {Quinn}, S.~N., {Vanderburg}, A., {et~al.} 2020, \apjl, 892,
  L7, \dodoi{10.3847/2041-8213/ab7302}

\bibitem[{{Jenkins} {et~al.}(2016){Jenkins}, {Twicken}, {McCauliff},
  {Campbell}, {Sanderfer}, {Lung}, {Mansouri-Samani}, {Girouard}, {Tenenbaum},
  {Klaus}, {Smith}, {Caldwell}, {Chacon}, {Henze}, {Heiges}, {Latham},
  {Morgan}, {Swade}, {Rinehart}, \& {Vanderspek}}]{jenkins2016spoc}
{Jenkins}, J.~M., {Twicken}, J.~D., {McCauliff}, S., {et~al.} 2016, in Society
  of Photo-Optical Instrumentation Engineers (SPIE) Conference Series, Vol.
  9913, Software and Cyberinfrastructure for Astronomy IV, ed. G.~{Chiozzi} \&
  J.~C. {Guzman}, 99133E, \dodoi{10.1117/12.2233418}

\bibitem[{{Kempton} {et~al.}(2018){Kempton}, {Bean}, {Louie}, {Deming}, {Koll},
  {Mansfield}, {Christiansen}, {L{\'o}pez-Morales}, {Swain}, {Zellem},
  {Ballard}, {Barclay}, {Barstow}, {Batalha}, {Beatty}, {Berta-Thompson},
  {Birkby}, {Buchhave}, {Charbonneau}, {Cowan}, {Crossfield}, {de Val-Borro},
  {Doyon}, {Dragomir}, {Gaidos}, {Heng}, {Hu}, {Kane}, {Kreidberg}, {Mallonn},
  {Morley}, {Narita}, {Nascimbeni}, {Pall{\'e}}, {Quintana}, {Rauscher},
  {Seager}, {Shkolnik}, {Sing}, {Sozzetti}, {Stassun}, {Valenti}, \& {von
  Essen}}]{kempton2018tsm}
{Kempton}, E. M.~R., {Bean}, J.~L., {Louie}, D.~R., {et~al.} 2018, \pasp, 130,
  114401, \dodoi{10.1088/1538-3873/aadf6f}

\bibitem[{{Kipping}(2013)}]{kipping2013ld}
{Kipping}, D.~M. 2013, \mnras, 435, 2152, \dodoi{10.1093/mnras/stt1435}

\bibitem[{{Knutson} {et~al.}(2014){Knutson}, {Fulton}, {Montet}, {Kao}, {Ngo},
  {Howard}, {Crepp}, {Hinkley}, {Bakos}, {Batygin}, {Johnson}, {Morton}, \&
  {Muirhead}}]{knutson2014hjfriendsI}
{Knutson}, H.~A., {Fulton}, B.~J., {Montet}, B.~T., {et~al.} 2014, \apj, 785,
  126, \dodoi{10.1088/0004-637X/785/2/126}

\bibitem[{{Kosiarek} \& {Crossfield}(2020)}]{kosiarek&crossfield2020}
{Kosiarek}, M.~R., \& {Crossfield}, I. J.~M. 2020, \aj, 159, 271,
  \dodoi{10.3847/1538-3881/ab8d3a}

\bibitem[{{Kov{\'a}cs} {et~al.}(2002){Kov{\'a}cs}, {Zucker}, \&
  {Mazeh}}]{kovacs2002bls}
{Kov{\'a}cs}, G., {Zucker}, S., \& {Mazeh}, T. 2002, \aap, 391, 369,
  \dodoi{10.1051/0004-6361:20020802}

\bibitem[{{Kozai}(1962)}]{kozai1962}
{Kozai}, Y. 1962, \aj, 67, 591, \dodoi{10.1086/108790}

\bibitem[{{Kreidberg}(2015)}]{kreidberg2015batman}
{Kreidberg}, L. 2015, \pasp, 127, 1161, \dodoi{10.1086/683602}

\bibitem[{{Leleu} {et~al.}(2021){Leleu}, {Alibert}, {Hara}, {Hooton}, {Wilson},
  {Robutel}, {Delisle}, {Laskar}, {Hoyer}, {Lovis}, {Bryant}, {Ducrot},
  {Cabrera}, {Delrez}, {Acton}, {Adibekyan}, {Allart}, {Allende Prieto},
  {Alonso}, {Alves}, {Anderson}, {Angerhausen}, {Anglada Escud{\'e}},
  {Asquier}, {Barrado}, {Barros}, {Baumjohann}, {Bayliss}, {Beck}, {Beck},
  {Bekkelien}, {Benz}, {Billot}, {Bonfanti}, {Bonfils}, {Bouchy}, {Bourrier},
  {Bou{\'e}}, {Brandeker}, {Broeg}, {Buder}, {Burdanov}, {Burleigh},
  {B{\'a}rczy}, {Cameron}, {Chamberlain}, {Charnoz}, {Cooke}, {Corral Van
  Damme}, {Correia}, {Cristiani}, {Damasso}, {Davies}, {Deleuil}, {Demangeon},
  {Demory}, {Di Marcantonio}, {Di Persio}, {Dumusque}, {Ehrenreich}, {Erikson},
  {Figueira}, {Fortier}, {Fossati}, {Fridlund}, {Futyan}, {Gandolfi},
  {Garc{\'\i}a Mu{\~n}oz}, {Garcia}, {Gill}, {Gillen}, {Gillon}, {Goad},
  {Gonz{\'a}lez Hern{\'a}ndez}, {Guedel}, {G{\"u}nther}, {Haldemann},
  {Henderson}, {Heng}, {Hogan}, {Isaak}, {Jehin}, {Jenkins}, {Jord{\'a}n},
  {Kiss}, {Kristiansen}, {Lam}, {Lavie}, {Lecavelier des Etangs}, {Lendl},
  {Lillo-Box}, {Lo Curto}, {Magrin}, {Martins}, {Maxted}, {McCormac}, {Mehner},
  {Micela}, {Molaro}, {Moyano}, {Murray}, {Nascimbeni}, {Nunes}, {Olofsson},
  {Osborn}, {Oshagh}, {Ottensamer}, {Pagano}, {Pall{\'e}}, {Pedersen}, {Pepe},
  {Persson}, {Peter}, {Piotto}, {Polenta}, {Pollacco}, {Poretti}, {Pozuelos},
  {Queloz}, {Ragazzoni}, {Rando}, {Ratti}, {Rauer}, {Raynard}, {Rebolo},
  {Reimers}, {Ribas}, {Santos}, {Scandariato}, {Schneider}, {Sebastian},
  {Sestovic}, {Simon}, {Smith}, {Sousa}, {Sozzetti}, {Steller}, {Su{\'a}rez
  Mascare{\~n}o}, {Szab{\'o}}, {S{\'e}gransan}, {Thomas}, {Thompson},
  {Tilbrook}, {Triaud}, {Turner}, {Udry}, {Van Grootel}, {Venus}, {Verrecchia},
  {Vines}, {Walton}, {West}, {Wheatley}, {Wolter}, \& {Zapatero
  Osorio}}]{leleu2021toi178}
{Leleu}, A., {Alibert}, Y., {Hara}, N.~C., {et~al.} 2021, \aap, 649, A26,
  \dodoi{10.1051/0004-6361/202039767}

\bibitem[{{Lidov}(1962)}]{lidov1962}
{Lidov}, M.~L. 1962, \planss, 9, 719, \dodoi{10.1016/0032-0633(62)90129-0}

\bibitem[{{Lillo-Box} {et~al.}(2020){Lillo-Box}, {Figueira}, {Leleu},
  {Acu{\~n}a}, {Faria}, {Hara}, {Santos}, {Correia}, {Robutel}, {Deleuil},
  {Barrado}, {Sousa}, {Bonfils}, {Mousis}, {Almenara}, {Astudillo-Defru},
  {Marcq}, {Udry}, {Lovis}, \& {Pepe}}]{lillobox2020lhs1140}
{Lillo-Box}, J., {Figueira}, P., {Leleu}, A., {et~al.} 2020, \aap, 642, A121,
  \dodoi{10.1051/0004-6361/202038922}

\bibitem[{{Lin} {et~al.}(1996){Lin}, {Bodenheimer}, \&
  {Richardson}}]{lin199651pegmigration}
{Lin}, D.~N.~C., {Bodenheimer}, P., \& {Richardson}, D.~C. 1996, \nat, 380,
  606, \dodoi{10.1038/380606a0}

\bibitem[{{Lopez}(2017)}]{lopez2017uspmodels}
{Lopez}, E.~D. 2017, \mnras, 472, 245, \dodoi{10.1093/mnras/stx1558}

\bibitem[{{McQuillan} {et~al.}(2013){McQuillan}, {Aigrain}, \&
  {Mazeh}}]{mcquillan2013stellarrot}
{McQuillan}, A., {Aigrain}, S., \& {Mazeh}, T. 2013, \mnras, 432, 1203,
  \dodoi{10.1093/mnras/stt536}

\bibitem[{{Mustill} {et~al.}(2015){Mustill}, {Davies}, \&
  {Johansen}}]{mustill2015hem}
{Mustill}, A.~J., {Davies}, M.~B., \& {Johansen}, A. 2015, \apj, 808, 14,
  \dodoi{10.1088/0004-637X/808/1/14}

\bibitem[{Neath \& Cavanaugh(2012)}]{neath2012bic}
Neath, A.~A., \& Cavanaugh, J.~E. 2012, WIREs Computational Statistics, 4, 199,
  \dodoi{https://doi.org/10.1002/wics.199}

\bibitem[{{Nelson} {et~al.}(2000){Nelson}, {Papaloizou}, {Masset}, \&
  {Kley}}]{nelson2000migration}
{Nelson}, R.~P., {Papaloizou}, J. C.~B., {Masset}, F., \& {Kley}, W. 2000,
  \mnras, 318, 18, \dodoi{10.1046/j.1365-8711.2000.03605.x}

\bibitem[{{Neveu-VanMalle} {et~al.}(2016){Neveu-VanMalle}, {Queloz},
  {Anderson}, {Brown}, {Collier Cameron}, {Delrez}, {D{\'\i}az}, {Gillon},
  {Hellier}, {Jehin}, {Lister}, {Pepe}, {Rojo}, {S{\'e}gransan}, {Triaud},
  {Turner}, \& {Udry}}]{neveuVanMalle2016wasp47Coralie}
{Neveu-VanMalle}, M., {Queloz}, D., {Anderson}, D.~R., {et~al.} 2016, \aap,
  586, A93, \dodoi{10.1051/0004-6361/201526965}

\bibitem[{{Osborn} \& {Bayliss}(2020)}]{osborn2020planetmetalcorr}
{Osborn}, A., \& {Bayliss}, D. 2020, \mnras, 491, 4481,
  \dodoi{10.1093/mnras/stz3207}

\bibitem[{{Osborn} {et~al.}(2021){Osborn}, {Armstrong}, {Adibekyan}, {Collins},
  {Delgado-Mena}, {Howell}, {Hellier}, {King}, {Lillo-Box}, {Nielsen}, {Otegi},
  {Santos}, {Ziegler}, {Anderson}, {Brice{\~n}o}, {Burke}, {Bayliss},
  {Barrado}, {Bryant}, {Brown}, {Barros}, {Bouchy}, {Caldwell}, {Conti},
  {D{\'\i}az}, {Dragomir}, {Deleuil}, {Demangeon}, {Dorn}, {Daylan},
  {Figueira}, {Helled}, {Hoyer}, {Jenkins}, {Jensen}, {Latham}, {Law}, {Louie},
  {Mann}, {Osborn}, {Pollacco}, {Rodriguez}, {Rackham}, {Ricker}, {Scott},
  {Sousa}, {Seager}, {Stassun}, {Smith}, {Str{\o}m}, {Udry}, {Villase{\~n}or},
  {Vanderspek}, {West}, {Wheatley}, \& {Winn}}]{osborn2021GPtoi755}
{Osborn}, H.~P., {Armstrong}, D.~J., {Adibekyan}, V., {et~al.} 2021, \mnras,
  502, 4842, \dodoi{10.1093/mnras/stab182}

\bibitem[{{Oshagh} {et~al.}(2017){Oshagh}, {Santos}, {Figueira}, {Barros},
  {Donati}, {Adibekyan}, {Faria}, {Watson}, {Cegla}, {Dumusque}, {H{\'e}brard},
  {Demangeon}, {Dreizler}, {Boisse}, {Deleuil}, {Bonfils}, {Pepe}, \&
  {Udry}}]{oshagh2017activityindicators}
{Oshagh}, M., {Santos}, N.~C., {Figueira}, P., {et~al.} 2017, \aap, 606, A107,
  \dodoi{10.1051/0004-6361/201731139}

\bibitem[{{Papaloizou} \& {Larwood}(2000)}]{ppaploizou2000discmigration}
{Papaloizou}, J.~C.~B., \& {Larwood}, J.~D. 2000, \mnras, 315, 823,
  \dodoi{10.1046/j.1365-8711.2000.03466.x}

\bibitem[{{Penz} {et~al.}(2008){Penz}, {Micela}, \&
  {Lammer}}]{penz2008exomassloss}
{Penz}, T., {Micela}, G., \& {Lammer}, H. 2008, \aap, 477, 309,
  \dodoi{10.1051/0004-6361:20078364}

\bibitem[{{Pepe} {et~al.}(2020){Pepe}, {Cristiani}, {Rebolo}, {Santos},
  {Dekker}, {Cabral}, {Di Marcantonio}, {Figueira}, {Lo Curto}, {Lovis},
  {Mayor}, {M{\'e}gevand}, {Molaro}, {Riva}, {Zapatero Osorio}, {Amate},
  {Manescau}, {Pasquini}, {Zerbi}, {Adibekyan}, {Abreu}, {Affolter}, {Alibert},
  {Aliverti}, {Allart}, {Allende Prieto}, {{\'A}lvarez}, {Alves}, {Avila},
  {Baldini}, {Bandy}, {Barros}, {Benz}, {Bianco}, {Borsa}, {Bourrier},
  {Bouchy}, {Broeg}, {Calderone}, {Cirami}, {Coelho}, {Conconi}, {Coretti},
  {Cumani}, {Cupani}, {D'Odorico}, {Damasso}, {Deiries}, {Delabre},
  {Demangeon}, {Dumusque}, {Ehrenreich}, {Faria}, {Fragoso}, {Genolet},
  {Genoni}, {G{\'e}nova Santos}, {Gonz{\'a}lez Hern{\'a}ndez}, {Hughes},
  {Iwert}, {Kerber}, {Knudstrup}, {Landoni}, {Lavie}, {Lillo-Box}, {Lizon},
  {Maire}, {Martins}, {Mehner}, {Micela}, {Modigliani}, {Monteiro}, {Monteiro},
  {Moschetti}, {Murphy}, {Nunes}, {Oggioni}, {Oliveira}, {Oshagh}, {Pall{\'e}},
  {Pariani}, {Poretti}, {Rasilla}, {Rebord{\~a}o}, {Redaelli}, {Santana
  Tschudi}, {Santin}, {Santos}, {S{\'e}gransan}, {Schmidt}, {Segovia},
  {Sosnowska}, {Sozzetti}, {Sousa}, {Span{\`o}}, {Su{\'a}rez Mascare{\~n}o},
  {Tabernero}, {Tenegi}, {Udry}, \& {Zanutta}}]{pepe2020espresso}
{Pepe}, F., {Cristiani}, S., {Rebolo}, R., {et~al.} 2020, arXiv e-prints,
  arXiv:2010.00316.
\newblock \doarXiv{2010.00316}

\bibitem[{{Poon} {et~al.}(2021){Poon}, {Nelson}, \&
  {Coleman}}]{poon2021insituformation}
{Poon}, S. T.~S., {Nelson}, R.~P., \& {Coleman}, G. A.~L. 2021, arXiv e-prints,
  arXiv:2105.08553.
\newblock \doarXiv{2105.08553}

\bibitem[{{Ricker} {et~al.}(2014){Ricker}, {Winn}, {Vanderspek}, {Latham},
  {Bakos}, {Bean}, {Berta-Thompson}, {Brown}, {Buchhave}, {Butler}, {Butler},
  {Chaplin}, {Charbonneau}, {Christensen-Dalsgaard}, {Clampin}, {Deming},
  {Doty}, {De Lee}, {Dressing}, {Dunham}, {Endl}, {Fressin}, {Ge}, {Henning},
  {Holman}, {Howard}, {Ida}, {Jenkins}, {Jernigan}, {Johnson}, {Kaltenegger},
  {Kawai}, {Kjeldsen}, {Laughlin}, {Levine}, {Lin}, {Lissauer}, {MacQueen},
  {Marcy}, {McCullough}, {Morton}, {Narita}, {Paegert}, {Palle}, {Pepe},
  {Pepper}, {Quirrenbach}, {Rinehart}, {Sasselov}, {Sato}, {Seager},
  {Sozzetti}, {Stassun}, {Sullivan}, {Szentgyorgyi}, {Torres}, {Udry}, \&
  {Villasenor}}]{ricker2014tess}
{Ricker}, G.~R., {Winn}, J.~N., {Vanderspek}, R., {et~al.} 2014, in Society of
  Photo-Optical Instrumentation Engineers (SPIE) Conference Series, Vol. 9143,
  Space Telescopes and Instrumentation 2014: Optical, Infrared, and Millimeter
  Wave, ed. J.~{Oschmann}, Jacobus~M., M.~{Clampin}, G.~G. {Fazio}, \& H.~A.
  {MacEwen}, 914320, \dodoi{10.1117/12.2063489}

\bibitem[{Salvatier {et~al.}(2016)Salvatier, Wiecki, \&
  Fonnesbeck}]{exoplanet:pymc3}
Salvatier, J., Wiecki, T.~V., \& Fonnesbeck, C. 2016, PeerJ Computer Science,
  2, e55

\bibitem[{{Sanchis-Ojeda} {et~al.}(2015){Sanchis-Ojeda}, {Winn}, {Dai},
  {Howard}, {Isaacson}, {Marcy}, {Petigura}, {Sinukoff}, {Weiss}, {Albrecht},
  {Hirano}, \& {Rogers}}]{sanchisojeda2015wasp47RM}
{Sanchis-Ojeda}, R., {Winn}, J.~N., {Dai}, F., {et~al.} 2015, \apjl, 812, L11,
  \dodoi{10.1088/2041-8205/812/1/L11}

\bibitem[{{Sanz-Forcada} {et~al.}(2011){Sanz-Forcada}, {Micela}, {Ribas},
  {Pollock}, {Eiroa}, {Velasco}, {Solano}, \&
  {Garc{\'\i}a-{\'A}lvarez}}]{sanzforcada2011photoevap}
{Sanz-Forcada}, J., {Micela}, G., {Ribas}, I., {et~al.} 2011, \aap, 532, A6,
  \dodoi{10.1051/0004-6361/201116594}

\bibitem[{{Schwarz}(1978)}]{schwarz1978bic}
{Schwarz}, G. 1978, Annals of Statistics, 6, 461

\bibitem[{{Shan} {et~al.}(2021){Shan}, {Yang}, {Lu}, {Wei}, {Tian}, {Zhang},
  {Guo}, {Cui}, {Yang}, {Zhang}, \& {Liu}}]{shan2021hotjupiterephems}
{Shan}, S.-S., {Yang}, F., {Lu}, Y.-J., {et~al.} 2021, arXiv e-prints,
  arXiv:2111.06678.
\newblock \doarXiv{2111.06678}

\bibitem[{{Sinukoff} {et~al.}(2017){Sinukoff}, {Howard}, {Petigura}, {Fulton},
  {Isaacson}, {Weiss}, {Brewer}, {Hansen}, {Hirsch}, {Christiansen}, {Crepp},
  {Crossfield}, {Schlieder}, {Ciardi}, {Beichman}, {Knutson}, {Benneke},
  {Dressing}, {Livingston}, {Deck}, {L{\'e}pine}, \&
  {Rogers}}]{sinukoff2017wasp47hires}
{Sinukoff}, E., {Howard}, A.~W., {Petigura}, E.~A., {et~al.} 2017, \aj, 153,
  70, \dodoi{10.3847/1538-3881/153/2/70}

\bibitem[{{Skrutskie} {et~al.}(2006){Skrutskie}, {Cutri}, {Stiening},
  {Weinberg}, {Schneider}, {Carpenter}, {Beichman}, {Capps}, {Chester},
  {Elias}, {Huchra}, {Liebert}, {Lonsdale}, {Monet}, {Price}, {Seitzer},
  {Jarrett}, {Kirkpatrick}, {Gizis}, {Howard}, {Evans}, {Fowler}, {Fullmer},
  {Hurt}, {Light}, {Kopan}, {Marsh}, {McCallon}, {Tam}, {Van Dyk}, \&
  {Wheelock}}]{skrutskie2006twomass}
{Skrutskie}, M.~F., {Cutri}, R.~M., {Stiening}, R., {et~al.} 2006, \aj, 131,
  1163, \dodoi{10.1086/498708}

\bibitem[{{Sozzetti} {et~al.}(2021){Sozzetti}, {Damasso}, {Bonomo}, {Alibert},
  {Sousa}, {Adibekyan}, {Zapatero Osorio}, {Gonz{\'a}lez Hern{\'a}ndez},
  {Barros}, {Lillo-Box}, {Stassun}, {Winn}, {Cristiani}, {Pepe}, {Rebolo},
  {Santos}, {Allart}, {Barclay}, {Bouchy}, {Cabral}, {Ciardi}, {Di
  Marcantonio}, {D'Odorico}, {Ehrenreich}, {Fasnaugh}, {Figueira}, {Haldemann},
  {Jenkins}, {Latham}, {Lavie}, {Lo Curto}, {Lovis}, {Martins}, {M{\'e}gevand},
  {Mehner}, {Micela}, {Molaro}, {Nunes}, {Oshagh}, {Otegi}, {Pall{\'e}},
  {Poretti}, {Ricker}, {Rodriguez}, {Seager}, {Su{\'a}rez Mascare{\~n}o},
  {Twicken}, \& {Udry}}]{sozzetti2021toi130}
{Sozzetti}, A., {Damasso}, M., {Bonomo}, A.~S., {et~al.} 2021, \aap, 648, A75,
  \dodoi{10.1051/0004-6361/202040034}

\bibitem[{{Stassun} {et~al.}(2019){Stassun}, {Oelkers}, {Paegert}, {Torres},
  {Pepper}, {De Lee}, {Collins}, {Latham}, {Muirhead}, {Chittidi},
  {Rojas-Ayala}, {Fleming}, {Rose}, {Tenenbaum}, {Ting}, {Kane}, {Barclay},
  {Bean}, {Brassuer}, {Charbonneau}, {Ge}, {Lissauer}, {Mann}, {McLean},
  {Mullally}, {Narita}, {Plavchan}, {Ricker}, {Sasselov}, {Seager}, {Sharma},
  {Shiao}, {Sozzetti}, {Stello}, {Vanderspek}, {Wallace}, \&
  {Winn}}]{stassun2019ticv8}
{Stassun}, K.~G., {Oelkers}, R.~J., {Paegert}, M., {et~al.} 2019, \aj, 158,
  138, \dodoi{10.3847/1538-3881/ab3467}

\bibitem[{{Steffen} {et~al.}(2012){Steffen}, {Ragozzine}, {Fabrycky}, {Carter},
  {Ford}, {Holman}, {Rowe}, {Welsh}, {Borucki}, {Boss}, {Ciardi}, \&
  {Quinn}}]{steffen2012}
{Steffen}, J.~H., {Ragozzine}, D., {Fabrycky}, D.~C., {et~al.} 2012,
  Proceedings of the National Academy of Science, 109, 7982,
  \dodoi{10.1073/pnas.1120970109}

\bibitem[{{Stevenson} {et~al.}(2016){Stevenson}, {Lewis}, {Bean}, {Beichman},
  {Fraine}, {Kilpatrick}, {Krick}, {Lothringer}, {Mandell}, {Valenti}, {Agol},
  {Angerhausen}, {Barstow}, {Birkmann}, {Burrows}, {Charbonneau}, {Cowan},
  {Crouzet}, {Cubillos}, {Curry}, {Dalba}, {de Wit}, {Deming}, {D{\'e}sert},
  {Doyon}, {Dragomir}, {Ehrenreich}, {Fortney}, {Garc{\'\i}a Mu{\~n}oz},
  {Gibson}, {Gizis}, {Greene}, {Harrington}, {Heng}, {Kataria}, {Kempton},
  {Knutson}, {Kreidberg}, {Lafreni{\`e}re}, {Lagage}, {Line}, {Lopez-Morales},
  {Madhusudhan}, {Morley}, {Rocchetto}, {Schlawin}, {Shkolnik}, {Shporer},
  {Sing}, {Todorov}, {Tucker}, \& {Wakeford}}]{stevenson2016jwstcomm}
{Stevenson}, K.~B., {Lewis}, N.~K., {Bean}, J.~L., {et~al.} 2016, \pasp, 128,
  094401, \dodoi{10.1088/1538-3873/128/967/094401}

\bibitem[{{Theano Development Team}(2016)}]{exoplanet:theano}
{Theano Development Team}. 2016, arXiv e-prints, abs/1605.02688.
\newblock \url{http://arxiv.org/abs/1605.02688}

\bibitem[{{Tsiaras} {et~al.}(2016){Tsiaras}, {Rocchetto}, {Waldmann}, {Venot},
  {Varley}, {Morello}, {Damiano}, {Tinetti}, {Barton}, {Yurchenko}, \&
  {Tennyson}}]{tsiaras201655cancrihstatmos}
{Tsiaras}, A., {Rocchetto}, M., {Waldmann}, I.~P., {et~al.} 2016, \apj, 820,
  99, \dodoi{10.3847/0004-637X/820/2/99}

\bibitem[{{Vanderburg} \& {Johnson}(2014)}]{vanderburg2014k2sff}
{Vanderburg}, A., \& {Johnson}, J.~A. 2014, \pasp, 126, 948,
  \dodoi{10.1086/678764}

\bibitem[{Vanderburg {et~al.}(2015)Vanderburg, Montet, Johnson, Buchhave, Zeng,
  Pepe, Cameron, Latham, Molinari, Udry, Lovis, Matthews, Cameron, Law, Bowler,
  Angus, Baranec, Bieryla, Boschin, Charbonneau, Cosentino, Dumusque, Figueira,
  Guenther, Harutyunyan, Hellier, Kuschnig, Lopez-Morales, Mayor, Micela,
  Moffat, Pedani, Phillips, Piotto, Pollacco, Queloz, Rice, Riddle, Rowe,
  Rucinski, Sasselov, S{\'{e}}gransan, Sozzetti, Szentgyorgyi, Watson, \&
  Weiss}]{vanderburg2015hip116454}
Vanderburg, A., Montet, B.~T., Johnson, J.~A., {et~al.} 2015, 800, 59,
  \dodoi{10.1088/0004-637x/800/1/59}

\bibitem[{{Vanderburg} {et~al.}(2017){Vanderburg}, {Becker}, {Buchhave},
  {Mortier}, {Lopez}, {Malavolta}, {Haywood}, {Latham}, {Charbonneau},
  {L{\'o}pez-Morales}, {Adams}, {Bonomo}, {Bouchy}, {Collier Cameron},
  {Cosentino}, {Di Fabrizio}, {Dumusque}, {Fiorenzano}, {Harutyunyan},
  {Johnson}, {Lorenzi}, {Lovis}, {Mayor}, {Micela}, {Molinari}, {Pedani},
  {Pepe}, {Piotto}, {Phillips}, {Rice}, {Sasselov}, {S{\'e}gransan},
  {Sozzetti}, {Udry}, \& {Watson}}]{vanderburg2017wasp47}
{Vanderburg}, A., {Becker}, J.~C., {Buchhave}, L.~A., {et~al.} 2017, \aj, 154,
  237, \dodoi{10.3847/1538-3881/aa918b}

\bibitem[{{Weiss} {et~al.}(2017){Weiss}, {Deck}, {Sinukoff}, {Petigura},
  {Agol}, {Lee}, {Becker}, {Howard}, {Isaacson}, {Crossfield}, {Fulton},
  {Hirsch}, \& {Benneke}}]{weiss2017wasp47rvttv}
{Weiss}, L.~M., {Deck}, K.~M., {Sinukoff}, E., {et~al.} 2017, \aj, 153, 265,
  \dodoi{10.3847/1538-3881/aa6c29}

\bibitem[{{Zeng} {et~al.}(2016){Zeng}, {Sasselov}, \&
  {Jacobsen}}]{zeng2016mrmodels}
{Zeng}, L., {Sasselov}, D.~D., \& {Jacobsen}, S.~B. 2016, \apj, 819, 127,
  \dodoi{10.3847/0004-637X/819/2/127}

\bibitem[{{Zeng} {et~al.}(2019){Zeng}, {Jacobsen}, {Sasselov}, {Petaev},
  {Vanderburg}, {Lopez-Morales}, {Perez-Mercader}, {Mattsson}, {Li}, {Heising},
  {Bonomo}, {Damasso}, {Berger}, {Cao}, {Levi}, \&
  {Wordsworth}}]{zeng2019mrmodels}
{Zeng}, L., {Jacobsen}, S.~B., {Sasselov}, D.~D., {et~al.} 2019, Proceedings of
  the National Academy of Science, 116, 9723, \dodoi{10.1073/pnas.1812905116}

\end{thebibliography}
\bibliographystyle{aasjournal}



\appendix
\section{ESPRESSO Radial-Velocities}
\begin{table*}[h!]
    \centering
    \caption{ESPRESSO radial-velocities of \Nstar.}
    \begin{tabular}{c|c|c|c|c|c|c}
        \hline
       Time  &  RV  & RV err  & CCF FWHM  & CCF FWHM err  & CCF Contrast & CCF Contrast err \\
        (BJD) &  kms$^{-1}$  &  kms$^{-1}$  &  kms$^{-1}$  &  kms$^{-1}$  &  \%  & \% \\
        \hline
    2458701.85502825 & -27.29026780 & 0.00041693 & 7.63531365 & 0.00083385 & 67.60791253 & 0.00738344 \\
    2458719.64853384 & -27.09946692 & 0.00040187 & 7.63677533 & 0.00080374 & 67.56452335 & 0.00711092 \\
    2458721.58500371 & -27.16635045 & 0.00049575 & 7.63476836 & 0.00099150 & 67.64179346 & 0.00878440 \\
    $^{\dagger}$2458725.61367832 & -27.13500443 & 0.00065487 & 7.62975492 & 0.00130973 & 67.63890353 & 0.01161098 \\
    $^{\dagger}$2458725.63068667 & -27.13984263 & 0.00062360 & 7.63615268 & 0.00124719 & 67.57960736 & 0.01103759 \\
    $^{\dagger}$2458725.64619095 & -27.14989021 & 0.00094470 & 7.64294765 & 0.00188939 & 67.57891409 & 0.01670599 \\
    2458727.61484710 & -27.16918161 & 0.00059743 & 7.63368748 & 0.00119485 & 67.62341354 & 0.01058467 \\
    2458727.67178015 & -27.15824555 & 0.00065289 & 7.63523000 & 0.00130578 & 67.62855131 & 0.01156585 \\
    2458727.73682302 & -27.14140987 & 0.00058531 & 7.63690880 & 0.00117061 & 67.61934501 & 0.01036494 \\
    2458728.78485747 & -27.00354831 & 0.00045830 & 7.63894339 & 0.00091660 & 67.57051680 & 0.00810783 \\
    2458729.56054875 & -27.09212322 & 0.00044374 & 7.63200749 & 0.00088747 & 67.59110555 & 0.00785971 \\
    2458729.61834544 & -27.10772602 & 0.00090117 & 7.63388129 & 0.00180233 & 67.68960271 & 0.01598127 \\
    2458736.67233204 & -27.03578795 & 0.00061549 & 7.62430696 & 0.00123098 & 67.32722849 & 0.01087029 \\
    $^{*}$2458740.75353437 & -27.05016571 & 0.00062895 & 7.57708547 & 0.00125790 & 65.82775520 & 0.01092832 \\
    $^{*}$2458740.77067861 & -27.04230379 & 0.00078360 & 7.54293260 & 0.00156719 & 64.96525397 & 0.01349782 \\
    2458742.61293777 & -27.21117435 & 0.00043227 & 7.63387461 & 0.00086453 & 67.34888424 & 0.00762724 \\
    $^{\dagger}$2458746.53807873 & -27.15209672 & 0.00044709 & 7.64715905 & 0.00089417 & 67.20359011 & 0.00785802 \\
    2458746.59477314 & -27.16215262 & 0.00044244 & 7.64505752 & 0.00088487 & 67.30934962 & 0.00779069 \\
    2458746.64130096 & -27.17045794 & 0.00043855 & 7.64196106 & 0.00087710 & 67.38690784 & 0.00773431 \\
    2458747.52008568 & -27.27233326 & 0.00050739 & 7.64561227 & 0.00101477 & 67.49901637 & 0.00895886 \\
    2458751.51960422 & -27.28267043 & 0.00048916 & 7.64547970 & 0.00097833 & 67.52486603 & 0.00864059 \\
    2458751.64509613 & -27.29039615 & 0.00045045 & 7.64359568 & 0.00090089 & 67.50749933 & 0.00795661 \\
    2458755.51389449 & -27.26383303 & 0.00048269 & 7.63960650 & 0.00096538 & 67.57048540 & 0.00853854 \\
    2458756.64036787 & -27.18625399 & 0.00044636 & 7.63636160 & 0.00089273 & 67.64158456 & 0.00790762 \\
    2458756.70863783 & -27.16868373 & 0.00045565 & 7.63589503 & 0.00091131 & 67.63068642 & 0.00807142 \\
    \hline
    \end{tabular}
    \begin{list}{}{}
        \item $\dagger$ Data points taken during a transit of \Nplanetb\ and so excluded from analysis
        \item * Data points excluded from analysis due to anomalous CCF profiles
    \end{list}
    \label{tab:espresso_rvs}
\end{table*}

\end{document}